\documentclass[notoc]{JHEP3}

\pdfoutput=1
\JHEPspecialurl{http://jhep.sissa.it/JOURNAL/JHEP3.tar.gz}

\usepackage{amsmath,epsfig}
\usepackage{bbm}
\usepackage{graphicx}

\newcommand\fverb{\setbox\fverbbox=\hbox\bgroup\verb}
\newcommand\fverbdo{\egroup\medskip\noindent%
			\fbox{\unhbox\fverbbox}\ }
\newcommand\fverbit{\egroup\item[\fbox{\unhbox\fverbbox}]}
\newbox\fverbbox

\newcommand{\etal}{{\it etal~}}
\newcommand{\gsim}{\lower.7ex\hbox{$\;\stackrel{\textstyle>}{\sim}\;$}}
\newcommand{\lsim}{\lower.7ex\hbox{$\;\stackrel{\textstyle<}{\sim}\;$}}

\title{Neutrino-Flavoured Sneutrino Dark Matter}

\author{John March-Russell, Christopher McCabe and Matthew McCullough \\
	Rudolf Peierls Centre for Theoretical Physics, University of Oxford, 1 Keble Road, Oxford, OX1 3NP, UK\\
	E-mail: \email{jmr@thphys.ox.ac.uk, mccabe@thphys.ox.ac.uk, mccull@thphys.ox.ac.uk}}

\preprint{OUTP-09 28 P}	% OR: \preprint{Aaaa/Mm/Yy\\Aaa-aa/Nnnnnn}
\abstract{A simple theory of supersymmetric dark matter (DM) naturally linked to neutrino flavour physics is studied.
The DM sector comprises a spectrum of mixed lhd-rhd sneutrino states where both the sneutrino flavour structure 
and mass splittings are determined by the associated neutrino masses and mixings.  Prospects for indirect detection from solar capture are good due to a large sneutrino-nucleon cross-section afforded by the inelastic splitting (solar capture limits exclude an explanation of DAMA/LIBRA).  We find parameter regions where all heavier states will have decayed, leaving only one flavour mixture of sneutrino as the candidate DM.  Such regions have a unique `smoking gun' signature---sneutrino annihilation in the Sun produces a pair of neutrino mass eigenstates free from vacuum oscillations, with the potential for detection at neutrino telescopes through the observation of a hard spectrum of $\nu_\mu$ and $\nu_\tau$ (for a normal neutrino hierarchy).  Next generation direct detection experiments can explore much of the parameter space through both elastic and inelastic scattering.  We show in detail that the observed neutrino masses and mixings can arise as a consequence of supersymmetry breaking effects in the sneutrino DM sector, consistent with all experimental constraints.   
}


\begin{document}

\section{Introduction}\label{intro}

The substantial observational evidence for non-baryonic dark matter (DM) is one of the clearest calls for new physics
beyond-the-Standard-Model, specifically the existence of an exotic stable or very-long-lived massive particle or
particles.   All this evidence, however, is indirect in form, and to-date does not fix the nature
of the DM particle beyond requiring it to be colour and electromagnetically neutral and relatively weakly self-interacting.
Despite this ignorance, one particular candidate has tended, until recently, to dominate theoretical speculation---the neutralino of the
minimal supersymmetric (SUSY) extension of the Standard Model, the MSSM.   Although the neutralino is the lightest supersymmetric
particle (LSP) over substantial regions of parameter space, and thus stable if $R$-parity conservation is assumed, the increasing
reach of both direct DM searches and collider constraints on the MSSM now pushes MSSM neutralino DM into uncomfortable
corners, as has been widely recognized.  The problems are two-fold: First, direct search experiments probe WIMP-nucleon cross sections orders of magnitude smaller than canonical weak scale cross sections; and second, given the direct search and collider constraints, the standard freeze-out mechanism of dark-matter-genesis generically generates a substantial excess of MSSM neutralino DM, limiting the allowed neutralino parameter space to special `fine-tuned' regions of parameter space with coannihilations or resonances.  Overall, neutralino DM matter works much less naturally than it did in the early 1990's
after the excitement of the LEP-I results pointing to SUSY unification.

An independent reason to possibly doubt the standard MSSM neutralino story is that it is based on an assumption
of simplicity of the dark sector that is probably not warranted given our experience of normal matter.  In the observable sector
there is an extraordinary richness of stable or very long-lived massive states, including more than one hundred essentially stable 
nuclear isotopes (on the timescale of the Hubble time), a stable charged lepton, and three essentially stable neutrino mass eigenstates
of various flavour compositions.  This is despite the fact that all global discrete and continuous symmetries such as individual lepton number, as well as total lepton and baryon numbers are violated (by at least neutrino mixing plus the electroweak anomaly, and likely GUT-suppressed interactions too which violate $B-L$ also).   There is no reason why we should expect the DM sector
to be any less complicated than the observed world, and in a sense the observation of massive neutrinos already directly supports
this hypothesis---the WIMP dark sector is already composed of at least a four-component cocktail of the three massive
neutrino species plus one more WIMP.   

Motivated by these arguments we, in this paper, investigate the physics of a very slight modification of the usual SUSY LSP
hypothesis that naturally and simply accommodates a much more structured and varied DM sector (and in fact one that
has attractive and potentially testable connections to neutrino flavour physics).  Specifically we study what we consider
to be one of the simplest alternatives to standard SUSY neutralino DM, though maintaining the SUSY framework that naturally
possesses the well-known successes of weak-scale SUSY theories, such as precision gauge coupling unification and a
solution to the hierarchy problem, namely, mixed lhd-rhd sneutrinos.  

Sneutrinos are potentially interesting alternate SUSY DM candidates since, unlike Majorana
neutralinos, they can carry flavour quantum numbers while being charge and colour neutral,
In addition they are the LSP in reasonably large regions of parameter space. \footnote{Another
attractive feature of considering such sneutrino DM is that it might afford a dynamical explanation 
of the baryon-to-dark-matter ratio $\Omega_b/\Omega_{DM} \simeq 1/5$ as the DM can now possess 
a lepton number asymmetry connected to the baryon-number-asymmetry, in contradistinction to Majorana
neutralino DM matter which can carry no such asymmetry.  We will not pursue these ideas in this work, but see
\cite{Kaplan:1991ah,Hooper:2004dc,Farrar:2004qy,Kitano:2004sv,Cosme:2005sb, Suematsu:2005zc,Abel:2006hr,
Gopalakrishna:2006kr,McDonald:2006if,Page:2007sh,Deppisch:2008bp,Kaplan:2009ag,Kribs:2009fy,Cohen:2009fz}
for studies along these lines.} The traditional reason
why sneutrinos have not been considered a good DM candidate is that pure lhd, ie, $SU(2)_L$-active,
sneutrinos have too large an annihilation cross section, and thus too small a freeze-out relic density
to be the observed DM.   However this problem is easily solved if one posits the existence of 
weak-scale rhd, ie, $SU(2)_L$-sterile, sneutrinos which mix with the lhd states once electroweak
symmetry and SUSY is broken.    Because of the observed family replication it is most natural to posit three rhd sneutrinos, one associated with each lepton flavour, in addition to the three standard
lhd sneutrinos.

Therefore we are led to a model with six weak-scale sneutrinos, that mix with each other and form, as we will argue
in detail, a structured DM sector carrying lepton flavour quantum numbers identical to that carried by the neutrino
mass eigenstates observed in neutrino oscillation experiments.   In addition, each of these complex scalars splits into
CP-odd and CP-even components, so the full spectrum of DM states is thus twelve real degrees of freedom with a variety
of masses.  

\begin{figure}[h]
\centering
\includegraphics[height=3.5in]{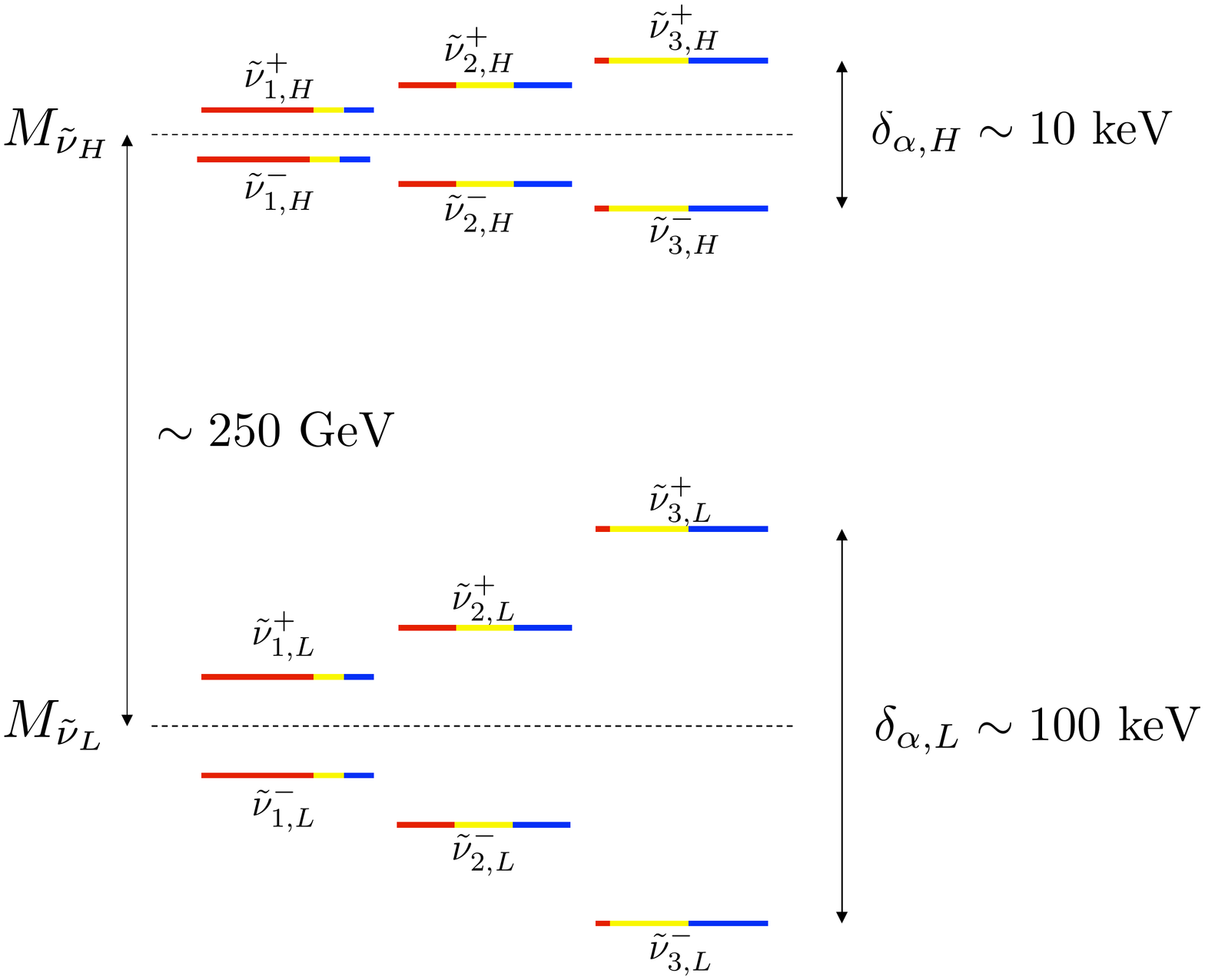}\caption{The spectrum of sneutrino masses after lifting of scalar and pseudoscalar mass degeneracy by the mass term $M_B^2$.  The red, yellow and blue lines represent, respectively, the electron, muon and tau lepton number flavour components, which mirrors the neutrino flavour structure (here we have assumed the normal neutrino mass hierarchy). Note that the 12 real degrees of freedom split into a heavy sector and a light sector of 6 states each, with further
fine structure splittings among these two groups, and the splittings between scalar and pseudoscalar components are proportional
to the neutrino masses.  This diagram is not to scale.}
\label{MassSpec}
\end{figure}

Shortening the story of the subsequent sections, what we find is a successful model comprising an entire sector of mixed sneutrino DM with the novel feature that the sneutrino flavour structure is identical to that of the neutrinos.  Further to this the sneutrinos have mass splittings between scalar and pseudoscalar components that are proportional to the neutrino masses. (See Figure \ref{MassSpec}.)
In addition to these interesting features we find regions of parameter space where the splittings are large enough that all states will have decayed down, leaving only one flavour of sneutrino as the candidate DM.  These regions of parameter space potentially have a unique `smoking gun' signature as pair annihilations of these sneutrinos through neutralino exchange produce just one neutrino mass eigenstate $\nu_3$ (for the normal neutrino mass hierarchy), free from vacuum oscillations, that has the potential for detection at future neutrino telescopes through the observation of a hard spectrum of $\nu_\mu$ and $\nu_\tau$ but not $\nu_e$.  These neutrinos would have energy equal to the sneutrino mass and could arise from annihilations in the Sun.

Since we require rhd sneutrinos at the weak scale our model also possesses sterile rhd neutrinos at this scale or below.   Naively one may therefore think that, as the rhd neutrinos are not superheavy, it is impossible to generate, eg, by the see-saw mechanism, an acceptable spectrum of light neutrinos as observed in neutrino oscillation experiments.  However as originally show by Arkani-Hamed \etal \cite{ArkaniHamed:2000bq, ArkaniHamed:2000kj} and Borzumati \etal \cite{Borzumati:2000mc, Borzumati:2000ya} and amplified on and extended by later authors (see, eg, \cite{TuckerSmith:2001hy, TuckerSmith:2004jv, MarchRussell:2004uf, Hambye:2004jf, West:2004me}) this is not the case.  In fact there occurs a new and extremely attractive mechanism of neutrino mass generation linked to supersymmetry breaking, in a sense generalizing the Giudice-Masiero mechanism for generating the $\mu$-term---the Higgsino
mass term---in the MSSM.  This model of neutrino masses allows for new explanations of the origin of the neutrino mass and flavour 
structure \cite{ArkaniHamed:2000bq, ArkaniHamed:2000kj,MarchRussell:2004uf,West:2004me} and elegantly accommodates such nice features as weak-scale resonant leptogenesis \cite{Hambye:2004jf}.  Since our main focus in this work is the novel sneutrino DM phenomenology we only very quickly summarize the physics of neutrino mass generation in as far as it impacts the sneutrino sector, and we strongly encourage the reader to refer to these other papers for more details.

Before we start, one other feature that deserves discussion is the fact that this class of models was the original and motivating
example of inelastic DM \cite{TuckerSmith:2001hy, TuckerSmith:2004jv}, providing a possible explanation of the
reported DAMA/LIBRA \cite{Bernabei:2008yi} signals consistent with the exclusions reported by other direct DM detection experiments.  
Since these original models involving sneutrinos, the idea of inelastic DM has been implemented in many more set ups, and
has been much studied (see, eg, \cite{Cui:2009xq} and references therein). This inelastic scattering also changes the phenomenology of DM solar capture and we include the exclusion limits from Refs. \cite{Nussinov:2009ft, Menon:2009qj} on the nucleon scattering cross-section.  These can be more constraining than the current generation of direct detection experiments over large regions of parameter space.

There are many previous studies of sneutrino DM, see eg, \cite{Hall:1997ah,Kolb:1999wx,Asaka:2005cn,Asaka:2006fs,Lee:2007mt,Arina:2007tm,Thomas:2007bu,Arina:2008yh,Arina:2008bb,Cerdeno:2008ep,Allahverdi:2009ae,Cerdeno:2009dv,Demir:2009kc,Allahverdi:2009se,Cerdeno:2009zz,Allahverdi:2009kr,Kumar:2009sf}, though most do not consider connecting the sneutrino and neutrino flavour structure. (For treatments of sneutrino DM taking limits from the generation of neutrino masses we refer the reader to \cite{Arina:2007tm,Thomas:2007bu,Kumar:2009sf}.)

The previous work most closely related to that considered here is Ref.\cite{Kumar:2009sf} where the flavour structure of the sneutrinos is also considered.  However our model has the novel feature that the neutrino and sneutrino flavour structures are exactly the same, allowing a predictive `smoking gun' signature for sneutrino annihilations into neutrinos, as discussed in Section~\ref{indirect_detection}.  Another important difference is that in our model there is an interplay between two terms contributing to the neutrino mass generation (see the off-diagonal entries of eqs.(\ref{massmat2})) that allow us to fit the observed neutrino masses consistent with successful sneutrino DM over a significant region of parameter space.
Finally, our analysis includes the important limits arising from solar capture, which excludes the possibility of an inelastic DM explanation of DAMA/LIBRA using mixed sneutrinos.  

In Section~\ref{model} we first discuss the specific model, outlining the generation of neutrino masses and the origin of the identical neutrino and sneutrino flavour structure, as well as the resulting mass spectrum of the sneutrino states. 
We then go on in Section~\ref{dark_matter}
to present the decay lifetimes for the different flavoured sneutrino states and consider the DM phenomenology of the sneutrinos, showing the regions of parameter space where they 
are good DM candidates.  In Section~\ref{detection} the potential unique signatures
of this model are considered, while our conclusions are contained in Section~\ref{conclusions}.
Two short appendices contain technical details.

\section{The Model}\label{model}

We start by summarizing the relevant field content and effective Lagrangian of the model we employ---that of
\cite{MarchRussell:2004uf}, to which we refer the reader for additional details.  

\subsection{Field content and interactions}\label{interactions}

To the field content of the MSSM we append
three $SU(3)\times SU(2)_L \times U(1)_Y$-sterile neutrinos and their sneutrino superpartners, $\tilde{n}_i$, which
combine into the chiral superfields $N_i$.  Here $i=1,2,3$ labels the generation.    The terms arising from the
superpotential are
\begin{equation}\label{efflag1}
\Delta\mathcal{L} = \int d^2 \theta  \left(\lambda_{ij} L_i N_j H_u + \frac{1}{2} M_N N_i N_i \right) .
\end{equation}
The SUSY-preserving superpotential term for the rhd neutrino mass, $M_N N_i N_i$, arises from a higher-dimension Kahler term involving a SUSY-breaking spurion $F$-component, in analogy with the Giudice-Masiero mechanism for the $\mu$-term, thus giving rhd neutrino masses at the weak scale $M_N \sim m_I^2/M \sim {\rm TeV}$. (Here $m_I\sim 10^{10}-10^{11}~{\rm GeV}$ is the intermediate scale at which SUSY is broken in the hidden sector, while $M$ is the reduced Planck scale.)    The Yukawa coupling $\lambda_{ij}$ between the lhd lepton doublet superfields, $L_i$, and the rhd neutrino supermultiplets, $N_i$, also arises from a Kahler term involving a spurion $F$-component and is suppressed in magnitude by a
factor of $|\lambda_{ij}| \sim m_I/M \sim 10^{-7} - 10^{-8}$.  These suppressions can be justified with an R-symmetry \cite{MarchRussell:2004uf}. 

In addition to the usual soft-SUSY breaking terms of the MSSM there are also SUSY-breaking terms of the form
\begin{equation}\label{efflag2}
\Delta\mathcal{L} =  m^2_{\tilde{n}}\, | \tilde{n}_i |^2 +
A\,\tilde{L}_i \tilde{n}_i h_u + \frac{1}{2}{\lambda_{ij} \, M^2_{B}}\, \tilde{n}_i \tilde{n}_j + h.c.~.
\end{equation}
Namely, TeV-scale, but flavour diagonal, soft mass and trilinear scalar $A$-terms, and a small
but significant rhd sneutrino lepton-number violating $B$-term with
coefficient $B^2_{ij}  \sim \lambda_{ij} M^2_{B} \sim m_I^2 (m_I/M)^3 \sim ({\rm TeV}^{5} / M)^{1/2} \sim (100 \text{ MeV})^2$ and flavour
structure identical to that of the neutrino Yukawa coupling, which we assume to be real.  Again this structure of suppressions and lepton flavour
breaking can be justified by the R-symmetry and flavour structure properties of the SUSY-breaking spurions.  More complicated
flavour breaking patterns, for instance those with non-diagonal $A$-terms, are also possible, but eq.(\ref{efflag1}) and (\ref{efflag2}) is the simplest structure that leads to a successful spectrum of masses and mixings in the neutrino sector (as we quickly recall in 
Section~\ref{nu_masses}).  It also leads to a simple and direct connection between the neutrino and sneutrino flavour structure
and masses.

\subsection{Sneutrino Masses}\label{sneumass}

In total we have six complex sneutrino fields, however these are split into scalar and pseudoscalar components by the sneutrino
$B$-term in eq.(\ref{efflag2}), leaving twelve mass eigenstates.  
After electroweak symmetry breaking the sneutrino mass matrix has the form:
\begin{equation}\label{massmat}
M^2_{ij} = \left( \begin{array}{cccc}
M^2_L \mathbbm{1}_3 & A v \sin \beta   \mathbbm{1}_3 & 0 & \lambda_{ij} M_N v \sin \beta \\
A v \sin \beta   \mathbbm{1}_3 & M^2_R  \mathbbm{1}_3 & \lambda_{ij} M_N v \sin \beta & \lambda_{ij} M^2_{B} \\
0 & \lambda_{ij} M_N v \sin \beta & M^2_L  \mathbbm{1}_3 & A v \sin \beta   \mathbbm{1}_3 \\
\lambda_{ij} M_N v \sin \beta & \lambda_{ij} M^2_{B} & A v \sin \beta   \mathbbm{1}_3 & M^2_R  \mathbbm{1}_3 \end{array} \right)
\end{equation}
where the basis is $(\tilde{\boldsymbol \nu}^*,\tilde{\boldsymbol n},\tilde{\boldsymbol \nu},\tilde{\boldsymbol n}^*)$. 
Due to the flavour diagonality of the unsuppressed terms after electroweak symmetry breaking, this $12\times12$ sneutrino mass matrix is made up of a total of sixteen $3\times3$ blocks which are either proportional to the identity matrix (from the A-term, Majorana mass F-term and D-terms), or $\lambda_{ij}$ (from the B-term and the rhd sneutrino F-term).
Therefore the sneutrino flavour structure is completely determined by the matrix that diagonalises $\lambda_{ij}$, which we will call $U_{\lambda}$.

After diagonalisation of the flavour structure the sneutrino mass matrix is in the form of sixteen diagonal $3\times3$ blocks, which by further rotations can be manipulated (detailed in Appendix \ref{A}) into the simple block-diagonal form of six $2\times2$ matrices, three for the scalars and pseudoscalars each:
\begin{align}\label{massmat2}
{M^2_{\tilde{\nu}^+_\alpha}} =  \left( \begin{array}{cc}
M^2_L & A v \sin \beta + \lambda_\alpha a_{\nu} M_B^2 \\
A v \sin \beta + \lambda_\alpha a_{\nu} M_B^2 & M_R^2 + \lambda_\alpha M_B^2 \end{array} \right)\\
{M^2_{\tilde{\nu}^-_\alpha}} =\left( \begin{array}{cc}
M^2_L & A v \sin \beta  - \lambda_\alpha a_{\nu} M_B^2 \\
A v \sin \beta - \lambda_\alpha a_{\nu} M_B^2 & M_R^2 - \lambda_\alpha M_B^2 \end{array} \right)
\end{align}
where $\tilde{\nu}^+$ and $\tilde{\nu}^-$ denote the scalar and pseudoscalar components respectively.  The subscript $\alpha = \{1,2,3\}$ denotes the mixed flavour sneutrino mass eigenstate (note the change from Latin to Greek indices on going from weak to mass eigenstates), $\lambda_\alpha$ are the eigenvalues of the $\lambda_{ij}$ matrix, $M^2_L = m^2_{\tilde{L}} + \frac{1}{2} M^2_Z \cos 2 \beta$ and $M^2_R = m^2_{\tilde{n}} + M^2_N$, where $m_{\tilde{L}}$ and $m_{\tilde{n}}$ are the usual soft mass terms, and, $v=174 \text{ GeV}$, is the Higgs expectation value.
We have also defined
\begin{equation} 
a_{\nu} = \frac{v M_N \sin \beta}{M_B^2}
\label{a_nu_param}
\end{equation}
which gives a measure of the relative size of the B-term and F-term contributions to the splitting of the CP-eigenstate masses.

These $2 \times 2$ matrices can then be diagonalised to complete the exact sneutrino rotation matrix $U$.  Further, in this form one can see there are twelve mass eigenstates: two $\tilde{\nu}^+$ and two $\tilde{\nu}^-$ for each flavour, and both the heavy and light $\tilde{\nu}^+$ and $\tilde{\nu}^-$ states for each flavour have their masses split due to $\lambda_{ij}$, becoming degenerate in the limit $\lambda_\alpha \rightarrow 0$.
Expanding to first order in the small parameter $\lambda_\alpha$ the splitting between the light $\tilde{\nu}^+_L$ and $\tilde{\nu}^-_L$ components for each flavour is given by:
\begin{equation}\label{splitting}
\delta_{\alpha,L} = {M_{\tilde{\nu}^+_{\alpha,L}}} - {M_{\tilde{\nu}^-_{\alpha,L}}} = \lambda_\alpha \frac{M_B^2}{M_{\tilde{\nu}_L}} (\cos^2 \theta-a_{\nu} \sin2\theta) \sim 100 \text{ keV}
\end{equation}
where $M_{\tilde{\nu}_L}$ and $\theta$ are the lightest mass eigenstate and the rotation angle which diagonalises the $2\times2$ matrices in eq.(\ref{massmat2}) in the limit $\lambda_\alpha \rightarrow 0$.  Similarly, the splitting between the $\tilde{\nu}^+_H$ and $\tilde{\nu}^-_H$ components is given by $\delta_{\alpha,H} = {M_{\tilde{\nu}^+_{\alpha,H}}} - {M_{\tilde{\nu}^-_{\alpha,H}}} = \lambda_\alpha \frac{M_B^2}{M_{\tilde{\nu}_H}} (\sin^2 \theta+a_{\nu} \sin2\theta) \sim 10 \text{ keV}$, where $M_{\tilde{\nu}_L}$ is the heavy mass eigenstate in the limit  $\lambda_\alpha \rightarrow 0$. Therefore we are left with a sneutrino mass spectrum as shown schematically in Figure~\ref{MassSpec}.

\subsection{Neutrino Masses}\label{nu_masses}

\begin{figure}[]
\centering
\includegraphics[height=1.0in]{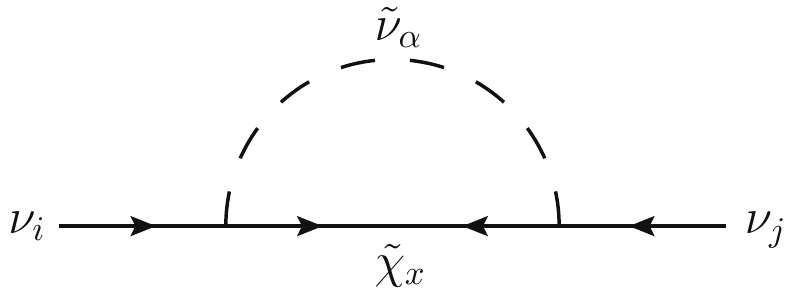}\caption{Feynman diagram corresponding to the radiative neutrino mass generation.}
\label{loopdiag}
\end{figure}

As the Majorana mass for the rhd neutrinos ($M_N \sim 1$ TeV) is very small compared to the standard seesaw set-up ($M_N \sim 10^{10} - 10^{15}$ GeV) one would naturally expect the seesaw contribution to the light neutrino masses to be large, however here this contribution is suppressed by the square of the small neutrino Yukawa coupling ($\lambda^2 \sim 10^{-14} - 10^{-16}$) and instead the dominant Majorana masses for the neutrinos arise radiatively through loops involving sneutrinos and neutralinos.  This is due to the combination of the lifting of the degeneracy between $\tilde{\nu}^+$ and $\tilde{\nu}^-$ sneutrino states and the Majorana nature of the neutralinos and can be seen schematically through loop diagrams involving perturbative mass-insertions, as in \cite{MarchRussell:2004uf}.  However, as it is possible to solve for the sneutrino masses and mixing matrices analytically, the exact one-loop contribution to the neutrino masses can be calculated from the Feynman diagram in Figure~\ref{loopdiag}.

Correspondingly, the neutrino masses are given by;
\begin{equation}\label{loopmass}
{M_{\nu_{ij}}} = \frac{1}{2} \left( \frac{M_Z}{4 \pi v} \right)^2 \sum^{12}_{\alpha=1} \sum^4_{x=1} m_{\chi_x} (N_{x1} \sin \theta_W - N_{x2} \cos \theta_W)^2  U_{i,\alpha}^{\dagger} L(r_\alpha,m_{\chi_x}) U_{\alpha,j+6}
\end{equation}
where $\alpha$ runs over the twelve sneutrino mass eigenstates, $N_{xy}$ are the neutralino mixing matrices, $m_{\chi_x}$ the neutralino masses, $r_\alpha = M_\alpha/m_{\chi_x}$ and:\footnote{As $\sum^{12}_{\alpha=1} U_{\alpha,i} {U^{\dagger}}_{j,\alpha} = \delta_{ij}$ the last two terms vanish when summing over the sneutrinos as they are independent of the sneutrino masses and $\alpha$.}
\begin{equation}\label{loopfact}
L(r,m_{\chi_x})= \frac{r^2 \ln(r^2)}{1-r^2} +1+\ln \left(\frac{\Lambda^2}{m_{\chi_x}^2}\right)
\end{equation}

The parameter dependence and flavour structure of the neutrino masses is more apparent by expanding the previous exact one-loop formula to first order in the $\lambda_{ij}$ matrix.  This gives;
\begin{equation}\label{pertmass}\begin{split}
{M_{\nu_{ij}}} &= \lambda_{ij} \left( \frac{M_Z M_B}{4 \pi v} \right)^2 \sum^4_{x=1} \frac{1}{m_{\chi_x}} (N_{x1} \sin \theta_W - N_{x2} \cos \theta_W)^2 \\
 &\quad \cdot \left(\sin^2\theta(\cos^2\theta-a_\nu \sin 2\theta)L_1(x_x,y_x)+\cos^2\theta(\sin^2\theta+a_\nu \sin 2\theta)L_2(x_x,y_x)\right)
 \end{split}\end{equation}
where $L_1$ and $L_2$, defined in Appendix \ref{B}, are the loop contributions from the sneutrino soft mass term and the F-term contributions.  The parameter $a_\nu$, defined in eq.(\ref{a_nu_param}), gives the freedom to generate neutrino masses that aren't too large by canceling these two contributions off each other.

From eq.(\ref{pertmass}) one can see that the neutrino masses are proportional to the eigenvalues of $\lambda_{ij}$.  Combining this with eq.(\ref{splitting}) it is clear thar $\delta_{\alpha} \propto M_{\nu_{\alpha}}$.  In order to obtain the correct neutrino masses it is necessary that $\lambda_\alpha/\lambda_\beta = M_{\nu_{\alpha}}/M_{\nu_{\beta}}$ and therefore $\delta_{\alpha}/\delta_{\beta} = M_{\nu_{\alpha}}/M_{\nu_{\beta}}$.  Once these ratios and all other parameters are set, then the overall magnitude of the neutrino masses depends on the parameter $a_\nu$.  We allow $a_\nu$ to take values $-1 < a_\nu < 1$, and find that for reasonable values of this parameter, neutrino masses below cosmological bounds and with $M_{\nu_{3}} > \sqrt{\Delta m^2_{12}+\Delta m^2_{23}}$ are obtained.  This will be discussed in more detail in Section~\ref{natural}.

It is clear from eq.(\ref{pertmass}), and is shown to all orders in $\lambda_{ij}$ in Appendix \ref{B}, that the flavour structure of the neutrino mass matrix takes a simple form. The PMNS neutrino mixing matrix is given simply by $U_{PMNS} \equiv U_\lambda$, where $U_\lambda$ was defined previously as the matrix which diagonalizes $\lambda_{ij}$, and also completely determines the flavour structure of the sneutrinos. Therefore the flavour structure of the neutrinos is automatically identical to that of the sneutrinos in this model. This enables us to make definite statements about possible `smoking gun' signatures.

Hence, in order to reproduce the measured neutrino mixing and mass parameters the form of $\lambda_{ij}$ is completely known, up to an overall normalisation factor which is absorbed into $M_B^2$. Since $\delta_{\alpha} \propto M_{\nu_{\alpha}}$, the largest sneutrino splitting $\delta_{\alpha}$ is in correspondence with the largest neutrino mass. For the normal neutrino mass hierarchy, we know the heaviest neutrino's flavour structure is almost exactly half $\mu$ and $\tau$, therefore we know the four sneutrino states $\tilde{\nu}_3$ (two heavy scalar and pseudoscalar, and two light scalar and pseudoscalar states) must also have this flavour structure. This is shown schematically in Figure \ref{MassSpec}.

\section{Dark Matter}\label{dark_matter}

In the previous section we have derived the sneutrino mass spectrum and shown that, as well as giving mass to the neutrinos, they also pass on their flavour structure to the neutrinos. In this section we would like to apply current experimental constraints to find the regions of parameter space where the flavoured sneutrinos are a good DM candidate.\footnote{For all of our studies we take the current total DM density to be given by the WMAP5+BAO+SN value $\Omega h^2=0.1143\pm0.0034$\cite{Komatsu:2008hk}.}
  
As explained above, the sneutrino spectrum is split into two groups of six particles with the same flavour structure as the neutrinos. In this section we will only be concerned with the possible transitions of the six lightest particles as the heavier ones very rapidly decay. 
We will find that there are three general possibilities for the current DM composition: 1) In a Hubble time all of the heavier sneutrinos could have decayed to the lightest state, $\tilde{\nu}_{3,L}^-$, which makes up all of the measured DM relic density; 2) The three lightest species survive to the present day, each making up a third of the measured relic density; 3) An intermediate situation where one or more of the heavier sneutrino states have decayed (or are currently decaying) down to the lighter states, but a non-trivial cocktail of different flavoured sneutrinos remains.
For pedagogical clarity, we shall assume in the following that the neutrinos follow the normal mass hierarchy, however we will briefly highlight the differences if the inverted mass hierarchy is assumed at the end of this section.

\subsection{Stable and meta-stable sneutrino spectrum and decays}\label{snu_decays}

\begin{figure}[h]
\centering
\includegraphics[height=2.5in]{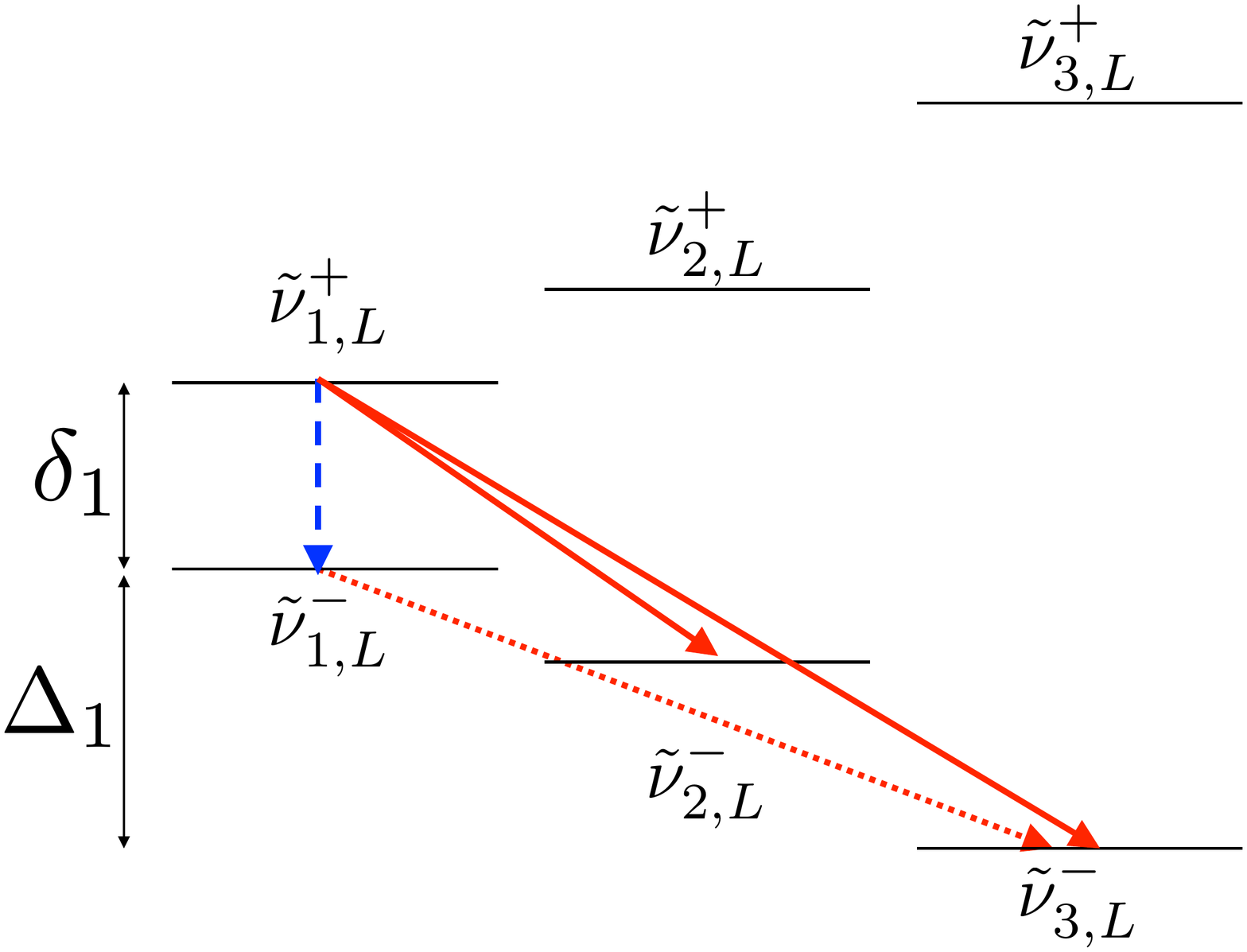}\caption{The six lightest sneutrino mass eigenstates (splittings not to scale). $\tilde{\nu}_{1,L}^+$ can decay to $\tilde{\nu}_{1,L}^-$ by $\nu\nu$ emission through intermediate Z exchange, as shown by the blue line, or to any of the lighter states by $\nu\nu$ emission through a neutralino (red lines). $\tilde{\nu}_{1,L}^-$ and $\tilde{\nu}_{2,L}^-$ can decay to $\tilde{\nu}_{3,L}^-$ through an intermediate neutralino exchange. For some regions of parameter space, these decays will be longer than the age of the Universe.}
\label{fig:decays}
\end{figure}

The lifetime of the lighter six states is a sensitive function of the splittings $\delta_1$ and $\Delta_1$, (defined in Figure~\ref{fig:decays}), which as shown in eq.(\ref{splitting}) depend on the ratio $a_{\nu}$. As we will show later, there are regions of the parameter space of $a_{\nu}$ where the lifetime of the decays of $\tilde{\nu}_{1,L}^-$ and $\tilde{\nu}_{2,L}^-$ to $\tilde{\nu}_{3,L}^-$ are longer than the age of the Universe. In this case, the DM is comprised of equal abundances of all three states $\tilde{\nu}_{1,L}^-$, $\tilde{\nu}_{2,L}^-$ and $\tilde{\nu}_{3,L}^-$, and is therefore overall flavour neutral.  However there are also portions of parameter space where the states will have decayed to only $\tilde{\nu}_{3,L}^-$, leading to one flavour of sneutrino as the DM candidate with potential signatures linked to the mixing properties of neutrinos.

Initially we consider the decay of the lightest scalar particle, $\tilde{\nu}_{1,L}^+$, as shown in Figure~\ref{fig:decays}. The particle can decay by $\nu \nu$ emission through intermediate Z exchange to the pseudoscalar particle of the same flavour, $\tilde{\nu}_{1,L}^-$, or by $\nu \nu$ emission through an intermediate neutralino exchange to any of the lighter pseudoscalar particles.  Typically the dominant decay mode is through the Z,
\begin{equation}
\Gamma_Z^{\tilde{\nu}_{1,L}^+} = \sin^4\theta \frac{G_F^2 \,\delta_1^5}{20\pi^3}\sim (9\times 10^4 \text{ yrs})^{-1}
\left(\frac{\sin\theta}{0.1}\right)^4\left(\frac{\delta_1}{100 \text{ keV}}\right)^5  \,  ,
\end{equation}
and is around an order of magnitude faster than neutralino-mediated decay.   The associated $\tilde{\nu}_{1,L}^+$ lifetime is, for all reasonable parameters, short enough that no $\tilde{\nu}_{\alpha,L}^+$ states survive to the current day.   Although these decays occur after Big Bang Nucleosynthesis there are no changes in the predicted light element abundances as the mass splittings are small and there is no hadronic energy injection.   There are similarly no constraints from spectral distortions in the CMB \cite{Hu:1993gc} from the energy injection due to Z-mediated decays to two photons, as the loop-suppressed branching fraction into photons is too small for these features to be observable.

We will now consider the decays of the pseudoscalar particles, and in particular, the decay of $\tilde{\nu}_{1,L}^-$ to $\tilde{\nu}_{3,L}^-$, as indicated by the dotted red arrow in Figure~\ref{fig:decays}, since it will be the fastest of the possible decays. If this lifetime is longer than the age of the universe, the lifetime for $\tilde{\nu}_{2,L}^-$ to decay to $\tilde{\nu}_{3,L}^-$ will also be longer. These decays can only occur by $\nu \nu$ emission through an intermediate neutralino. The decay width for typical parameters is:
\begin{align}
\Gamma_\chi^{\tilde{\nu}_{1,L}^-} &= \chi \sin^4\theta \left(\frac{M_z}{v}\right)^4\frac{\Delta_1^5}{960 \pi^3 m_{\chi_1}^2 m_{\tilde{\nu}}^2}
\\ &\sim(10^{10} \text{ yrs})^{-1} \left( \frac{\chi}{0.22} \right) \left( \frac{\sin \theta}{0.1} \right)^4 \left( \frac{\Delta_1}{25 \text{ keV}} \right)^5 \left( \frac{100 \text{ GeV}}{m_{\tilde{\nu}}} \right)^2 \left( \frac{200 \text{ GeV}}{m_{\chi_1}} \right)^2 
\label{light_decay}
\end{align}
where $ \chi= \left( \sum_x \frac{m_{\chi_1}}{m_{\chi_x}}  (N_{x1}\sin \theta_W-N_{x2}\cos\theta_W)^2  \right)^2+\left(  \sum_x \frac{m_{\chi_1} m_{\tilde{\nu}}}{m_{\chi_x}^2} |N_{x1}\sin \theta_W-N_{x2}\cos\theta_W|^2 \right)^2$ depends on the
neutralino mixing matrix, $N_{xy}$, and neutralino masses, $m_{\chi_x}$.

The decay width eq.(\ref{light_decay}) is in the range where, depending upon parameters, the pseudoscalar $\tilde{\nu}_{1,L}^-$ states can either have essentially all decayed by the present epoch, or can be substantially still present, or can be currently decaying.   For the purposes of matching to the observed DM density, however, this ambiguity is immaterial as each heavier
state decays to one only marginally lighter state.  The same is true for the decays of the $\tilde{\nu}_{\alpha,L}^+$ states, so  in computing the total freeze-out relic density it is a good approximation to sum the contribution of all six light states ignoring the
later effect of decays.

\subsection{DM relic density and experimental constraints}\label{relic}

\begin{figure}[hhh]
\centering
\includegraphics[height=1.8in]{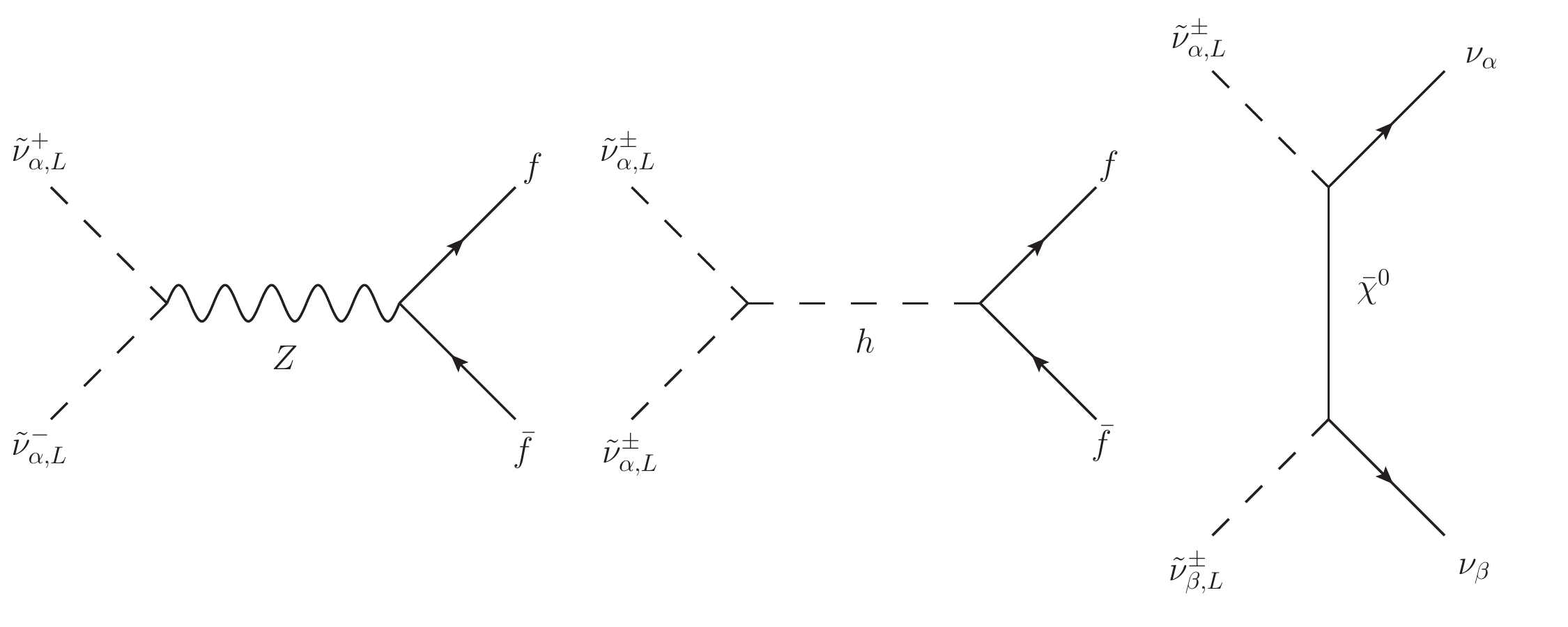}\caption{The dominant processes which set the sneutrino relic density.}
\label{fig:relic}
\end{figure}

Although various thermal and non-thermal mechanisms for generating the observed DM relic density are possible, we assume that the relic density is generated by the standard thermal freeze-out process.   In particular we do not here consider the possibility of using the calculable and IR-dominated thermal freeze-in process recently advocated by Hall \etal \cite{Hall:2009bx}, as this would 
apply for supersymmetric theories with Dirac neutrino masses, rather than the model outlined in Section~\ref{model}.

\begin{figure}[ht]
\centering
\includegraphics[height=3.1in]{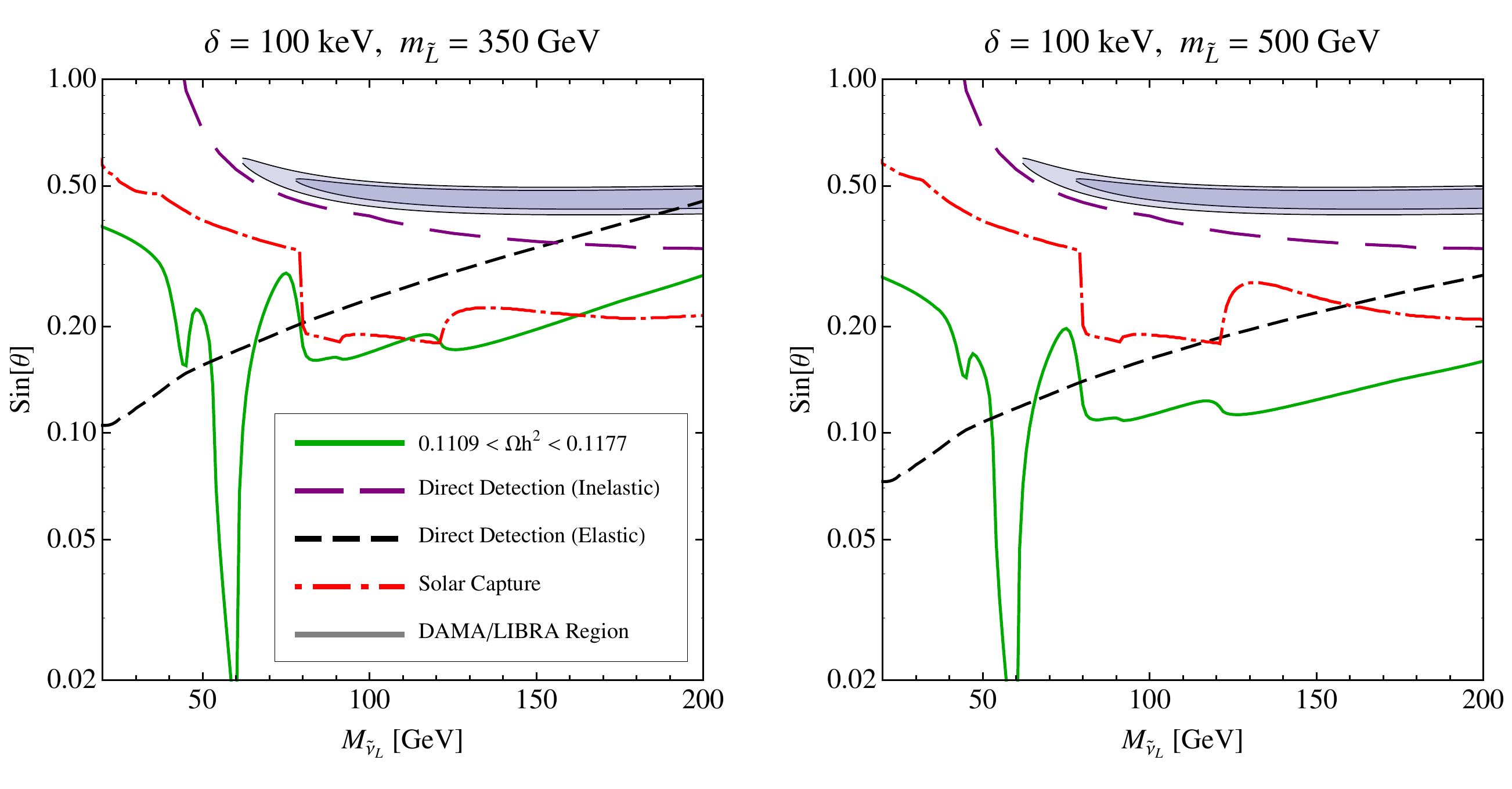}
\caption{Exclusion curves and thermal relic density constraints in the $\sin\theta$ - $M_{\tilde{\nu}_L}$ plane for $M_1=\mu=250 \text{ GeV}$, $M_2 = 500 \text{ GeV}$ and $\tan\beta = 10$, and with Higgs mass 121 GeV. The green solid line indicates the value of $\sin\theta$ where the observed DM relic density is generated thermally. The dashed lines give the exclusion curves from direct detection experiments and the shaded regions show the values of $\sin\theta$ which fit the DAMA/LIBRA experiment at 90\% and 99\% confidence levels. The dot-dashed curve gives the exclusion curve from the measurement of solar neutrinos from Super-Kamiokande. The green solid line below the dashed and dash-dotted lines indicate parameters where the sneutrinos are a good thermal DM candidate.  The two features below 60 GeV on this thermal relic density curve correspond to the Higgs- and Z-funnel regions.  The Z-funnel is a co-annihilation effect and so would be absent from
indirect detection signals.  Note that, with the exception of the Higgs funnel region, the elastic direct detection, and solar capture indirect detection limits are close to the region of parameter space required for a successful thermally generated relic density.}
\label{Limits}
\end{figure}

\begin{figure}[ht]
\centering
\includegraphics[height=3.1in]{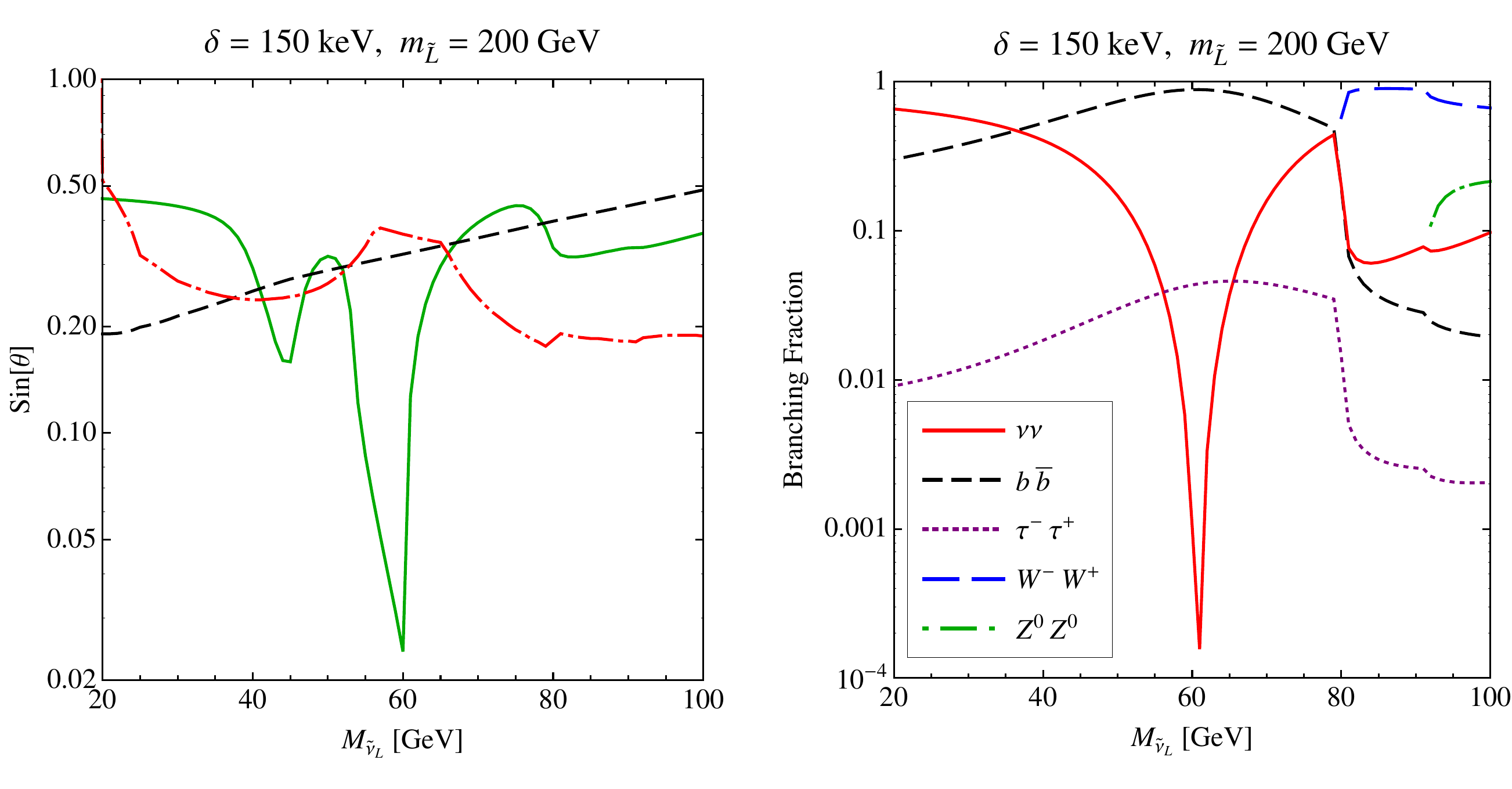}
\caption{Left panel: Exclusion curves and thermal relic density constraints for $M_1=\mu=150 \text{ GeV}$, $M_2 = 300 \text{ GeV}$ and $\tan\beta = 10$. As well as the Higgs-funnel region at $\sim 60$ GeV, the Z-funnel region at $\sim 45$ GeV is now also viable.
Again, note that, with the exception of the Higgs funnel region, the elastic direct detection, and solar capture indirect detection limits are close to the region of parameter space required for a successful thermally generated relic density.
Right panel: The branching fractions for the same parameters as in the left panel. For each value of $M_{\tilde{\nu}_L}$, we have chosen $\sin\theta$ to give the correct thermal relic density. At low masses the branching fractions to neutrinos dominates, and at masses just below the $W$ mass is substantial.  At masses above the $W$ mass, that to $W^+ W^-$ dominates.}
\label{ml200&bf}
\end{figure}

\begin{figure}[t]
\centering
\includegraphics[height=3.1in]{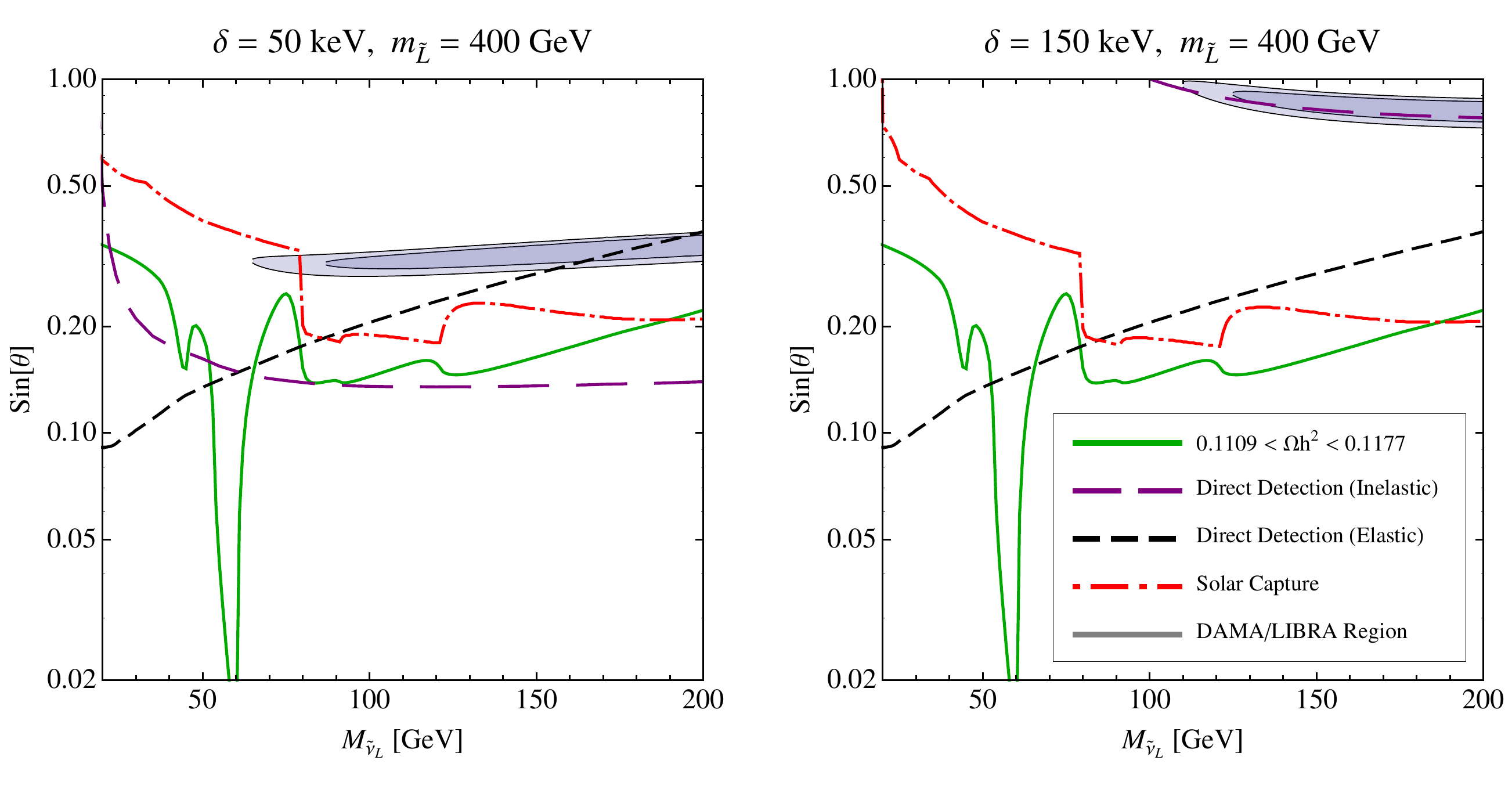}
\caption{The effect of changing $\delta$ while keeping everything else fixed. The relic density and elastic direct detection curves remain the same, while there is a small change at low masses in the solar capture limits. The major change is in the inelastic direct detection limits which change considerably due to the different kinematics of the collision with the target nucleus.}
\label{VaryDe}
\end{figure}

The three main processes which contribute to the freeze-out relic density calculation are shown in Figure~\ref{fig:relic}. We performed the calculation using micrOMEGAs 2.2 \cite{Belanger:2008sj} with model files created using LanHEP 3.0.4 \cite{Semenov:2008jy} and the results are shown as the solid green curves in Figures \ref{Limits}, \ref{ml200&bf} and \ref{VaryDe}.  Coannihilation between the six lightest sneutrinos, which is obviously important since they are almost degenerate in mass, is automatically included in micrOMEGAs. 
The dashed black and purple lines show the limits from the current generation of direct detection experiments on the DM-nucleon scattering cross section, or equivalently from using eqs.(\ref{Zcrosssection}) and (\ref{hcrosssection}), on the mixing angle $\sin\theta$. An interesting feature of this model is the possibility of two signals at future direct detection experiments; the DM can up-scatter inelastically via Z exchange, for example from $\tilde{\nu}_{3,L}^-$ to $\tilde{\nu}_{3,L}^+$, and therefore is sensitive to the small mass splitting $\delta_{\alpha,L}$, or can elastically scatter by Higgs exchange. This also means that we can constrain the parameter space in two regions due to the different kinematics of each collision. The elastic cross section is larger at smaller masses $M_{\tilde{\nu}_L}$, while the inelastic cross section limit is stronger for smaller $\delta_{\alpha, L}$, as we will explicitly demonstrate later. The cross section for coherent scattering off a nucleus at zero momentum transfer by Z exchange is 
\begin{equation}\label{Zcrosssection}
\sigma_N^Z= \sin^4\theta \frac{G_F^2\,\mu^2}{2 \pi}  \left((A-Z)-(1-4\sin^2\theta_W)Z\right)^2
\end{equation}
\noindent Here $\mu$ is the reduced mass for the sneutrino-nucleus, $Z$ the proton number and $A$, the number of protons and neutrons in the nucleus. The cross section for elastic scattering off a nucleon by Higgs exchange in the decoupling limit is given by:
\begin{equation}\label{hcrosssection}
\sigma_n^h= \frac{g_{hnn}^2}{8 \pi m_h^4}\left( \frac{m_n}{m_n+m_{\tilde{\nu}}}\right)^2 \left( A \sin\beta \sin 2\theta-\frac{M_Z^2}{v}\cos2\beta \sin^2\theta\right)^2
\end{equation}
\noindent where $m_{n}$ is the nucleon mass and $g_{hnn}=\frac{m_n}{\sqrt{2}v}\left( \sum_q^{u,d,s}f_{T_q}+\frac{2}{27}\sum_Q^{c,b,t}f_{T_G}\right)$. We use the values of $f_{T_q}$ and $f_{T_G}$ from Ref.\cite{Arina:2007tm} which give $g_{hnn}=1.43 \times 10^{-3}$.
 
The red dot-dashed lines show the indirect detection limits on $\sin\theta$ from the DM capture, and subsequent annihilation into neutrinos, by the Sun. At the masses we consider, the limits are set from the observation of solar neutrinos by Super-Kamiokande \cite{Desai:2004pq}. Note that the solar capture and annihilation limits are the most constraining limits over a sizeable region of parameter space.

In the limit $\lambda_\alpha \rightarrow 0$, the sneutrino masses and mixings are determined by $M_L$, $M_R$, $A$ and $\tan\beta$, however we find it more intuitive to trade in $M_L$, $M_R$ and $A$ for $M_{\tilde{L}}$, $M_{\tilde{\nu}_L}$ and $\theta$. Furthermore, once $a_{\nu}$ and one of the $\delta_{\alpha, L}$ are specified, and $M1$, $M2$ and $\mu$ are chosen so that the neutralino masses $m_{\chi_x}$ and mixing angles $N_{xy}$ can be computed, the absolute neutrino masses can be calculated from eq.(\ref{pertmass}) and the other two $\delta_{\alpha, L}$ can be found from the ratio $\delta_{\alpha, L}/\delta_{\beta, L} = M_{\nu_{\alpha}}/M_{\nu_{\beta}}$. At the top of each figure we have displayed the value of  $m_{\tilde{L}}$ and $\delta$ we have fixed. Note that $\delta$ always refers to the smallest $\delta_{\alpha, L}$ from the sneutrinos which have not decayed.

For all further calculations we fix the MSSM parameters at the weak scale. We set $\tan\beta=10$, the pseudoscalar Higgs mass $M_A=500\text{ GeV}$, the rhd slepton soft mass $m_{\tilde{E}}=250\text{ GeV}$, the top soft coupling $A_t=-1\text{ TeV}$ and all other soft parameters to $1\text{ TeV}$ giving a Higgs mass of $121\text{ GeV}$. $M1$, $M2$ and $\mu$ are varied in different plots but we maintain the relation $M2=2 M1$.

The dominant channel which contributes to the relic density is through the Higgs s-channel because of the large A-terms. An especially noticeable feature in the relic density curve in Figures~\ref{Limits}, \ref{ml200&bf} and \ref{VaryDe} is the Higgs funnel at half the Higgs mass $\sim 60$ GeV. A similar effect can be seen at $\sim 45$ GeV, the Z funnel, but is much smaller because it is suppressed by $\sin^4\theta$ rather than $\sin^2 2\theta$, cf. eqs.(\ref{Zcrosssection}) and (\ref{hcrosssection}). The annihilation through the neutralino t-channel is important at lower mass ($<45$ GeV) and just below the $W$ mass, where, from the right panel of Figure~\ref{ml200&bf}, we see that the branching fraction into neutrinos can be large.

As $m_{\tilde{L}}$ increases with $\theta$ and $M_{\tilde{\nu}_L}$ held constant, the A-term increases, so smaller values of $\sin\theta$ are required to ensure the sneutrinos do not over-annihilate producing a relic density below the measured value. From eq.(\ref{hcrosssection}), we can also see that this increase in A is the reason why the elastic direct detection limits become stronger as $m_{\tilde{L}}$ is increased.

For $m_{\tilde{L}}<350$ GeV, the only region of parameter space not excluded from elastic direct detection and solar limits is below the $W$ mass; as well as the Higgs-funnel region, the left panel of Figure \ref{ml200&bf} demonstrates that the Z-funnel region at $\sim45$ GeV also opens up. For $m_{\tilde{L}}>500$ GeV, the elastic direct detection limits and relic density curve move to lower values of $\sin\theta$ together so that no new lower values of $m_{\tilde{\nu}_L}$ are allowed. 

Changing $\delta$ has virtually no effect on the relic density calculation and the elastic direct detection limits, while the change in the solar limits is very minor. However as Figure \ref{VaryDe} shows, the major difference is in the inelastic direct detection limits which change considerably.

\begin{figure}[hhh]
\centering
\includegraphics[height=3.0in]{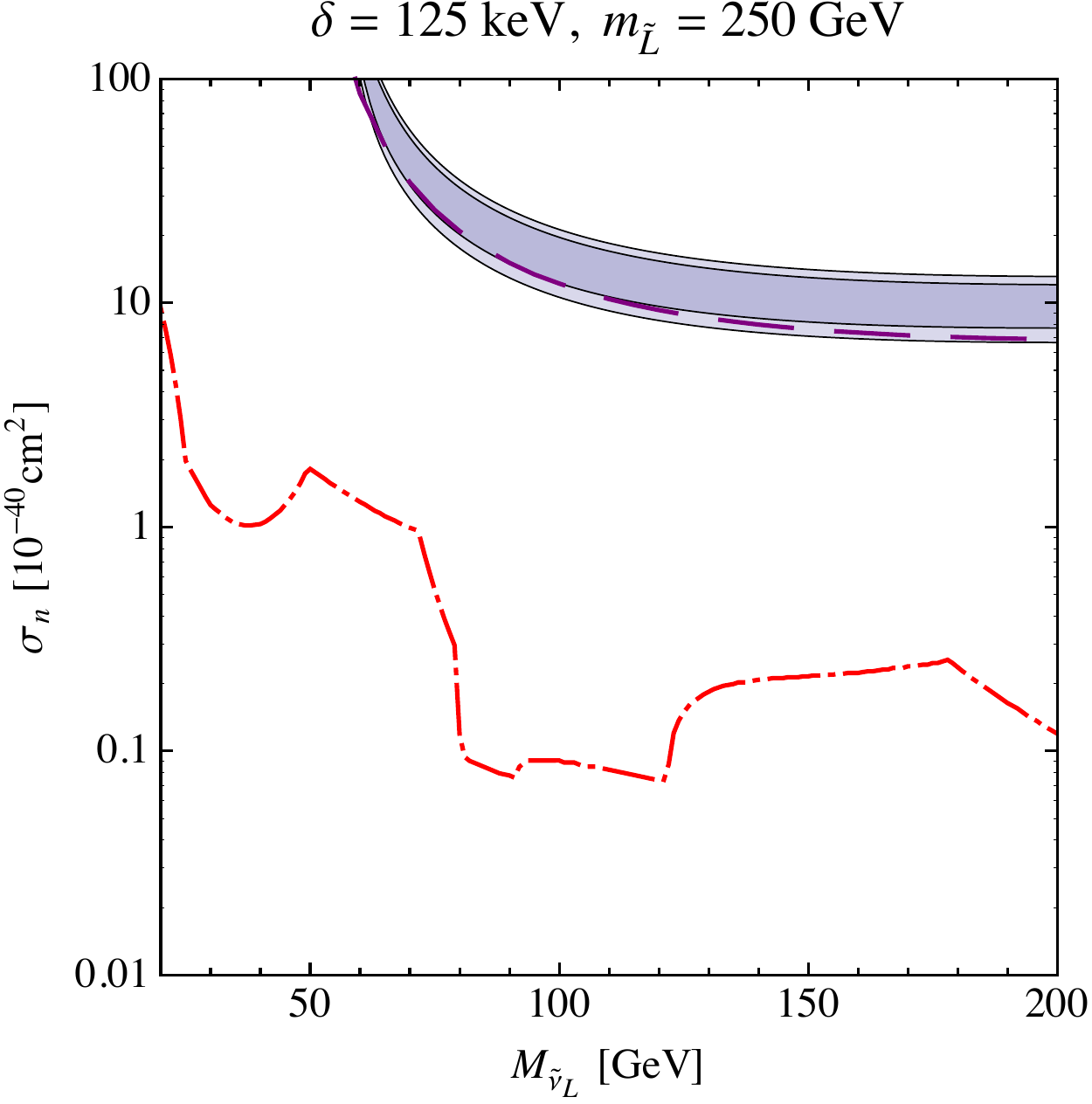}
\caption{The grey shaded regions show the values of the scattering cross section $\sigma_n$ which fit the DAMA/LIBRA experiment at 90\% and 99\% confidence levels. Also shown are the best direct detection limits (purple dashed) and solar capture limits (red dot dashed). Clearly the solar capture limits completely exclude the DAMA/LIBRA region in this model.}
\label{fig:DAMA}
\end{figure}

The somewhat erratic appearance of the solar limits in Figures \ref{Limits}, \ref{ml200&bf}, \ref{VaryDe} and \ref{fig:DAMA} is because we have only plotted the strongest constraint arising from the different possible annihilation channels; $\nu\nu$, $\tau\tau$, $b \bar{b}$, $W^+W^-$, $ZZ$ and $hh$. Typically we found that for $M_{\tilde{\nu}_L}<M_W$, the limits from annihilation into $\nu\nu$ dominate except for when $M_{\tilde{\nu}_L}\sim M_h/2$,  where the $\tau\tau$ limits dominate. This is because the neutrinos are produced through t-channel neutralino exchange which are not enhanced in the Higgs funnel, as other massive particles are. For $M_{\tilde{\nu}_L}>M_W$, the $W^+W^-$ limits typically dominate. The branching fractions were calculated using micrOMEGAs and we used the limits on the annihilation branching fraction derived using the methods in \cite{Nussinov:2009ft}. We have also only shown the strongest limits from the direct detection experiments. We assume the standard halo model and refer the reader to Ref. \cite{MarchRussell:2008dy} for details on how these limits were calculated and the experimental details.

Since our model naturally includes a splitting of order $100\text{ keV}$, it is natural to ask whether this is consistent with the inelastic DM \cite{TuckerSmith:2001hy} explanation of the positive annual modulation signal measured by the DAMA/LIBRA \cite{Bernabei:2008yi} experiment and the null results from the other direct detection experiments (notably CDMS II \cite{Ahmed:2008eu} and XENON10 \cite{Angle:2007uj}). Although DAMA/LIBRA may still be consistent with the other direct detection experiments at $\delta \approx 125\text{ keV}$ \cite{Chang:2008gd, MarchRussell:2008dy, Cui:2009xq, Cline:2009xd, SchmidtHoberg:2009gn}, we find that when the solar capture limits from Refs. \cite{Nussinov:2009ft, Menon:2009qj} are included as in Figure~\ref{fig:DAMA}, the DAMA/LIBRA region is conclusively excluded in this model.

%\begin{figure}[hhh]
%\centering
%\includegraphics[height=3.8in]{branching.pdf}
%\caption{Annihilation branching ratios at $p_{CM} = 1 \text{ GeV}$ for a sneutrino-Higgs soft coupling of $A = 50 \text{GeV}$.  These branching ratios are important for the detection of neutrinos produced from annihilation in the sun.}
%\label{Branch}
%\end{figure}

\subsection{The allowed $a_\nu$ parameter space}\label{natural}

For a given set of the parameters $\{m_{\tilde{L}},\mu,M1,M2,\tan\beta \}$ one can see from Figures  \ref{Limits}, \ref{ml200&bf} and \ref{VaryDe} that the allowed values of $\sin\theta$ and $M_{\tilde{\nu}_L}$ are tightly constrained by limits from direct and indirect detection.  Although non-thermal mechanisms for generating a relic density are possible, we also impose that the correct relic density is generated thermally and this reduces the allowed parameter space further.  Therefore once $M_{\tilde{\nu}_L}$ is chosen, $\sin\theta$ is set by this constraint.

Finally it remains to constrain the parameter $a_\nu$ which contributes to the inelastic splittings and the neutrino masses through eqs. \ref{splitting} and \ref{pertmass}.  For a given splitting $\delta_{3,L}$ the allowed parameter space for $a_\nu$ is determined by the requirement that the neutrino masses satisfy both the cosmological bound, $\sum m_\nu < 0.61 \text{ eV}$ \cite{Strumia:2006db}, and the mass bound $M_{\nu_{3}} > \sqrt{\Delta m^2_{12}+\Delta m^2_{23}}$.
\begin{figure}[h]
\centering
\includegraphics[height=3.0in]{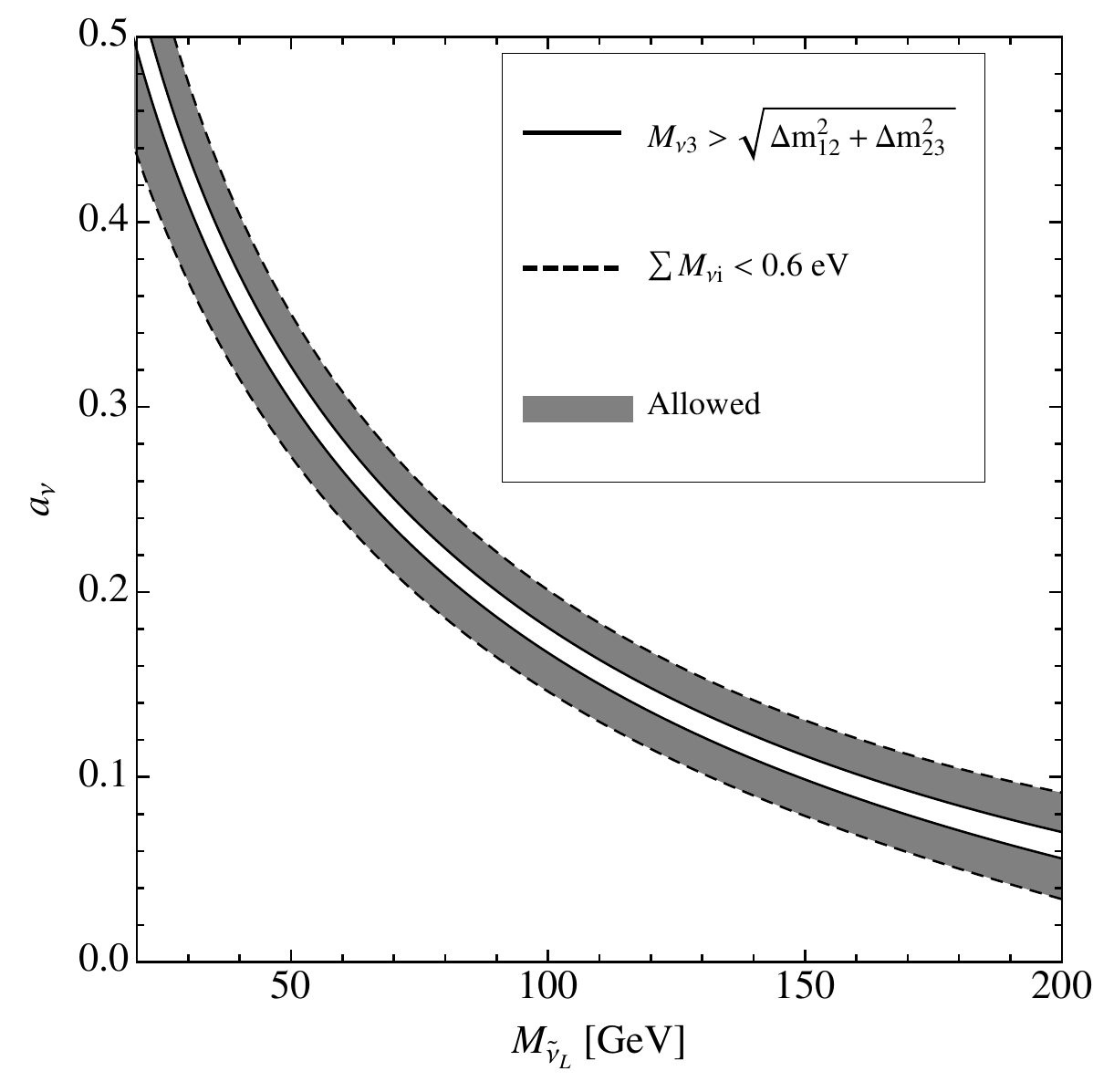}
\caption{Allowed regions of parameter space for the $a_\nu$ parameter for $\mu = M1 = 250 \text{GeV}$, $M2 = 500 \text{GeV}$, $m_{\tilde{\nu}_L} = 350$ GeV, $\sin \theta = 0.2$, $\tan \beta = 10$ and $\delta = 100$ keV.}
\label{aallow}
\end{figure}

In Figure \ref{aallow} we show the allowed values of $a_\nu$ for a given parameter set with varying WIMP mass $M_{\tilde{\nu}_L}$.  For a typical set of soft mass and electroweak parameters it is in general possible to find a value of $a_\nu$ for which the neutrino masses satisfy the bounds, although the allowed region may become small.

We have shown that within this model it is possible to generate all of the observed neutrino parameters and satisfy cosmological mass-sum bounds.  Concurrently the mixed sneutrinos have the same flavour structure as the neutrinos and are good DM candidates capable of providing the observed relic density and avoiding current direct and indirect detection limits.

It is now interesting to consider properties of the sneutrinos that could lead to a positive identification at future experiments.  A potentially unambiguous annihilation signal could arise if the DM sneutrinos have all decayed down to the lowest state.

\begin{figure}[h]
\centering
\includegraphics[height=3.0in]{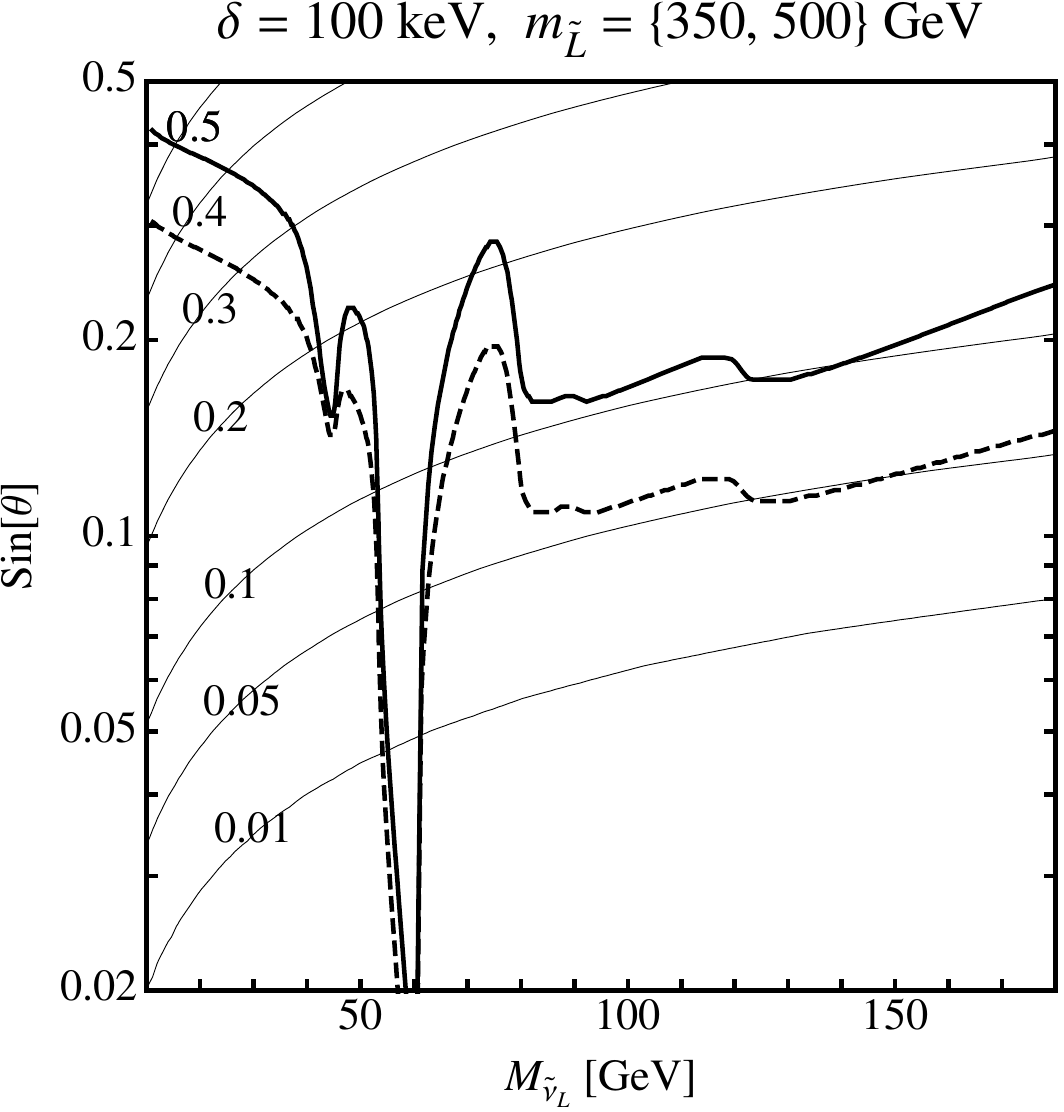}\hspace{0.2in}\includegraphics[height=3.0in]{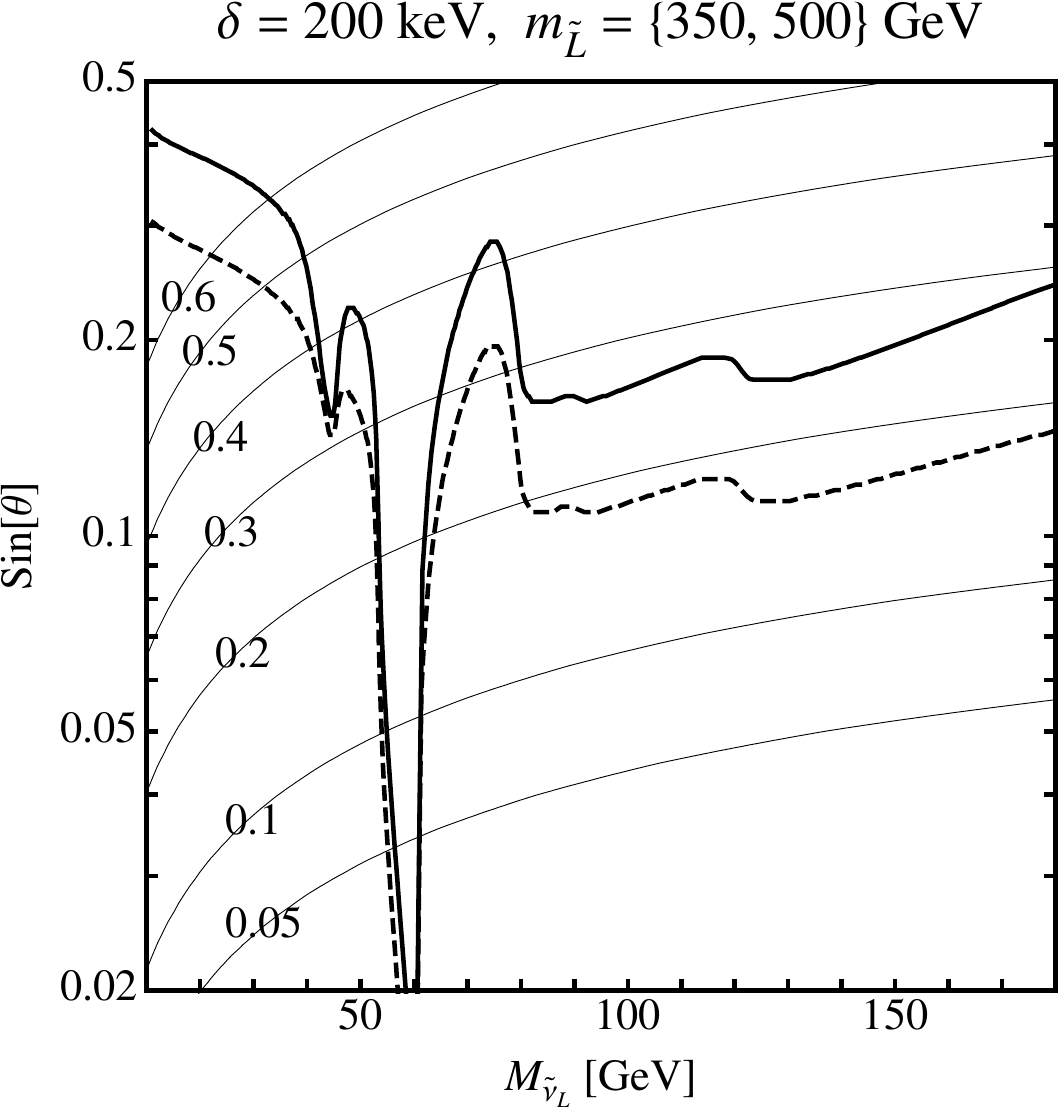}
\caption{Contours of constant fraction of $a_\nu$ parameter space for which the sneutrinos have decayed down to the lightest state (i.e. constant fraction of allowed values of $a_\nu$ for which $\tau_{\tilde{\nu}^{-}_{2,L}\rightarrow \tilde{\nu}^{-}_{3,L}} < 10^{10}$ yrs).  For comparison the regions where the observed relic density is produced are superimposed.  The SUSY-breaking parameters are $\mu = M1 = 250 \text{ GeV}$, $M2 = 500 \text{ GeV}$, $\tan \beta = 10$ and the solid and dashed lines are for $m_{\tilde{L}} = 350$ and $500$ GeV.}
\label{decayed}
\end{figure}

By choosing the same parameter set as that used to generate Figure \ref{Limits} and varying values of $\sin \theta$ and WIMP mass, we have calculated the fraction of the allowed parameter space of the parameter $a_\nu$ for which the sneutrinos will all have decayed down to the lightest state.  The results for two different values of $\delta$ are shown in Figure \ref{decayed}.  From this it seems that, for this model of DM, the fraction of parameter space for which the DM sneutrinos are made up of one state lies in the range $5-30\%$ for WIMP masses above $M_W$ and the range $0-45\%$ for masses below $M_W$.

\subsection{Neutrino mass hierarchy}\label{hierarchy}
This model does not have a preference for the neutrino mass hierarchy, since it is equally capable of generating either. In the discussion above we have assumed the normal mass hierarchy. Here we highlight how the phenomenology changes if the inverted hierarchy is assumed. Since $\delta_{\alpha,L} \propto M_{\nu_{\alpha}}$, the $\delta_{\alpha,L}$ follow the neutrino mass hierarchy so that in the normal hierarchy, $\delta_{3,L}$ is the largest splitting, and is much bigger than $\delta_{2,L}$ and $\delta_{1,L}$ which are expected to always be close together because $\delta_{2,L}^2-\delta_{1,L}^2 \propto \Delta m_{21}^2$, which is known to be small from solar neutrino measurements. Conversely in the inverted hierarchy, $\delta_{3,L}$ will be the smallest splitting, $\delta_{2,L}$ will be the largest, but $\delta_{1,L}$ will be similar because $\Delta m_{21}^2$ is small. Therefore  even if $\tilde{\nu}_{3,L}^-$ has decayed, we expect to have a significant fraction of both $\tilde{\nu}_{1,L}^-$ and $\tilde{\nu}_{2,L}^-$ and the overall flavour structure of the DM will depend precisely on the relative abundances of $\tilde{\nu}_{1,L}^-$ and $\tilde{\nu}_{2,L}^-$.

\section{Detection and Identification}\label{detection}

\subsection{Indirect Detection}\label{indirect_detection}

\begin{figure}[hh]
\centering
\includegraphics[height=2.8in]{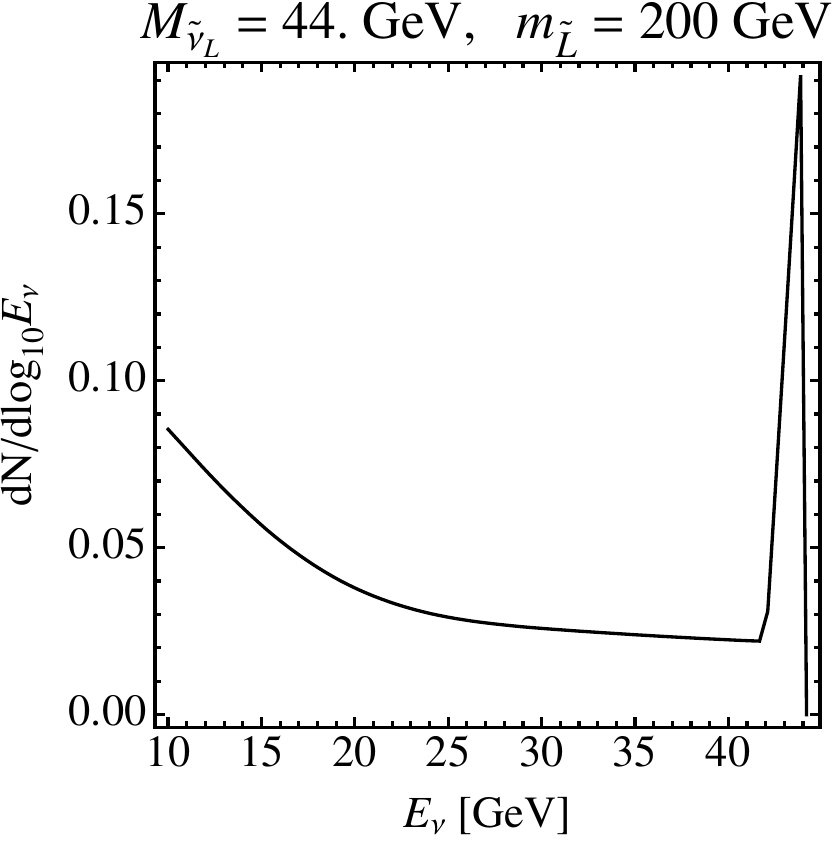}\hspace{0.1in}
\includegraphics[height=2.8in]{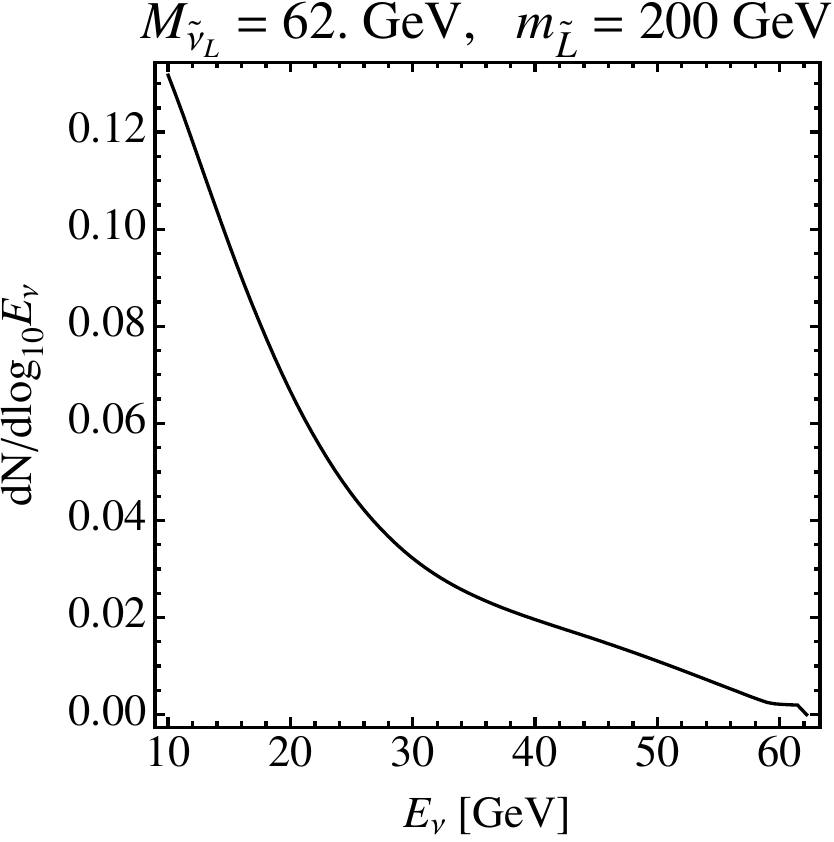}\vspace{0.1in}
\includegraphics[height=2.8in]{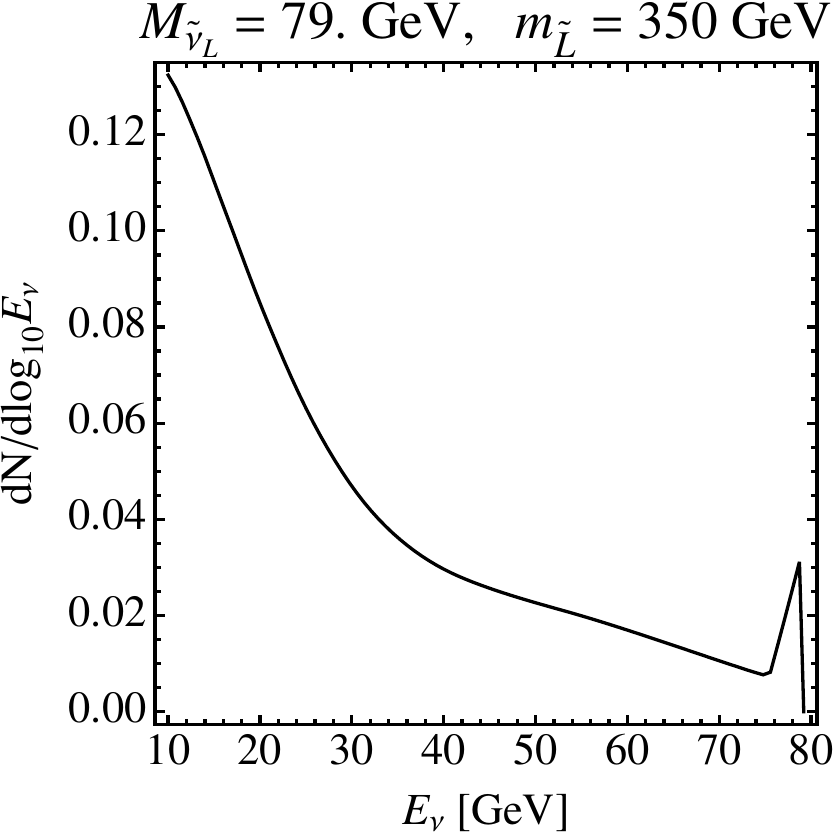}\hspace{0.1in}
\includegraphics[height=2.8in]{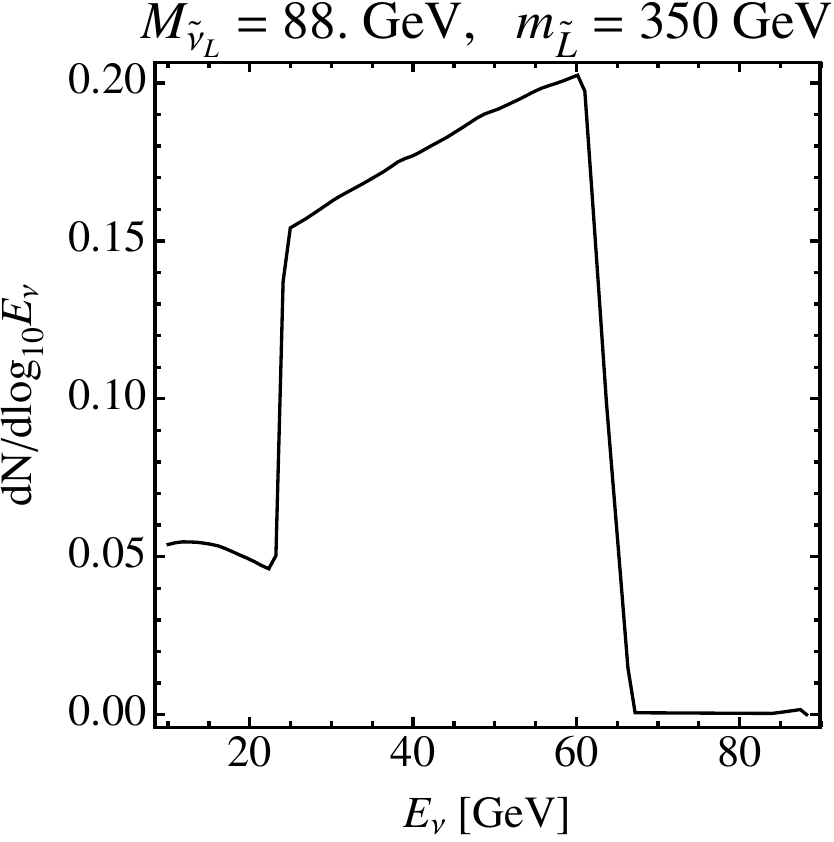}
\caption{The spectrum of neutrinos from the Sun per sneutrino annihilation for two sets of parameters lying just below current indirect detection bounds.  The SUSY-breaking parameters are as in Figure 6 and $m_{\tilde{L}}$ is varied in order to stay just below current indirect detection bounds for higher sneutrino masses.  The neutrino spectrum is evaluated using the branching fractions calculated using micrOMEGAs 2.2 \cite{Belanger:2008sj} along with the oscillation and propagation code publicly available \cite{Cirelli:2005gh}. 
For the two lhd panels the hard feature corresponds to two-body annihilation and gives neutrinos at energy $M_{{\tilde \nu}_L}$.  The branching fraction to neutrinos is large away from the Higgs funnel and below the $W$ mass, as shown in the rhd panel of Figure 6.  Note
that these are raw spectra.  In the rhd panels we illustrate cases where the direct 2-neutrino final state branching fraction is small, the
spectrum of neutrinos being dominated by secondary decays of annihilation products.}
\label{spectrum}
\end{figure}

Mixed sneutrinos present an interesting scenario for indirect detection through neutrino annihilation products for three reasons:
\begin{itemize}
\item  The inelastic splitting allows for a comparatively large WIMP-nucleon scattering cross section, $\sigma_n$, due to the kinematic suppression of scattering at direct detection experiments.  However, the large escape velocity of the Sun implies a typical WIMP kinetic energy in the Sun greater than this inelastic splitting, thus allowing a relatively large solar capture rate.
\item  Sneutrino annihilation can proceed via t-channel neutralino exchange to two neutrinos which propagate relatively unhindered (with some attenuation in the Sun) to the Earth. These neutrinos could be detected in future neutrino telescopes as a hard spectrum centered at the sneutrino mass, although this hard spectrum may lie below the flux of neutrinos from annihilations to other SM particles.  This is illustrated in Figure \ref{spectrum}.
\item  If the DM were made up of all three lowest lying states then annihilations to neutrinos would result in equal numbers of $\nu_e$, $\nu_\mu$ and $\nu_\tau$, with the directional flux of hard $\nu_\mu$ equalling roughly one third of the total flux of hard neutrinos.  If the sneutrinos have all decayed down to the lightest state then the hard spectrum of neutrinos would be comprised solely of the $\nu_3$ mass eigenstate, and the $\nu_\mu$ flux would equal roughly one half of the total flux for the normal hierarchy neutrino structure.  In the vacuum none of the neutrinos would oscillate as they are produced in the mass eigenstates $\nu_{1,2,3}$.  Moreover, if annihilations are purely to $\nu_3$ states in the sun, MSW effects would be negligible due to the small $\nu_e$ component of $\nu_3$.
\end{itemize}
As the WIMP-nucleon scattering cross section is typically large, sneutrino capture and annihilation in astrophysical bodies provides the most promising scenario for indirect detection.  However it should be noted that neutrino signals from annihilation in the galactic halo are not subject to the large uncertainties relating to the DM structure in the center of the galaxy \cite{Yuksel:2007ac}, and a dedicated analysis by neutrino telescope collaborations could place interesting limits on thermal relic WIMPs from annihilations resulting in final state neutrinos.

Due to the suppressed capture rate in the Earth from the inelastic splitting, we will only discuss potential signals from sneutrino capture in the Sun.  Although a thorough treatment of limits from upcoming neutrino telescopes is best left until after the collection of data, it is interesting to speculate about the detection potential.  Compared to existing Super-Kamiokande limits \cite{Desai:2004pq} projections of the performance of the completed IceCube + DeepCore detector indicate an order of magnitude increase in sensitivity to $\nu_\mu$ from the sun for WIMP masses $\sim 100$ GeV \cite{Halzen:2009vu}.  An order of magnitude increase in sensitivity to $\sigma_n$ corresponds to an increase in the limits on $\sin \theta$ by a factor $\sim 1.8$.  As the regions where the observed relic density is generated are already close to current indirect detection limits then inspection of Figures \ref{Limits}, \ref{ml200&bf} and \ref{VaryDe} would suggest that the completed IceCube detector will be sensitive to large portions of parameter space of this model for $m_{\tilde{L}} < 500$ GeV.

In Figure \ref{spectrum} we plot the neutrino spectrum at the Earth for various sneutrino masses where current direct and indirect detection bounds are just satisfied.  Inspecting Figure \ref{ml200&bf} one can see that when $M_{\tilde{\nu}_L}>M_W$ (respectively $M_Z$) then $W$ (and $Z$) bosons are the dominant annihilation products, while for $M_{\tilde{\nu}_L} \sim M_h/2$ then $\overline{b} b$ dominate due to the Higgs resonance.  Neutrinos from both these annihilations swamp the hard spectrum of neutrinos from neutralino exchange.  However when the sneutrino masses lie out of these ranges the sharp peak from annihilation to neutrinos can be seen where the annihilation to neutrinos is at its greatest.  Observation of such a feature would constitute a compelling case for sneutrino DM as direct annihilation of neutralinos to neutrinos is forbidden due to their Majorana nature.  As described in Ref. \cite{Allahverdi:2009ae}, the peak from annihilations to neutrinos would lead to a linear component in the muon energy spectrum at IceCube and, if large enough, could be distinguished from annihilation to other standard model particles which give a non-linear muon energy spectrum.  Further, if the flavour content of this component or alternatively just that of the highest energy neutrinos could be determined and was found to be non-democratic, then this would be an indication of neutrino-flavoured sneutrino DM.

We emphasize that if the sneutrinos have all decayed down to the lightest state then the prediction of this model for the ratios of the flavour content of the hard neutrino spectrum from annihilations through neutralinos is $\sim(0,1/2,1/2)$, in the basis ($f_e,f_\mu,f_\tau$),
exactly matching the flavour content of the heaviest neutrino in the normal hierarchy.
If the inverted hierarchy is assumed and the sneutrinos have decayed down to the two lightest states, which are very long lived due to their small mass splitting, the flavour content would be expected to take the form $\sim(1,1/2,1/2)$.

\subsection{Direct detection}
Referring the reader back to Figures \ref{Limits}, \ref{ml200&bf} and \ref{VaryDe} one can see that current direct detection bounds are close to large portions of parameter space for this model, whether for elastic scattering via Higgs exchange or inelastic scattering when the splitting $\delta$ is smaller than $\sim 100$ keV.  Therefore the prospects for the direct detection of sneutrino DM are good as future direct detection experiments such as EURECA \cite{Kraus:2007zz}, XENON100 \cite{Aprile:2009yh} and LUX \cite{kastens-2009-80} are expected to have significant increases in detection sensitivity.

Regardless of the size of the inelastic splitting the elastic scattering via Higgs exchange is unavoidable and future direct detection experiments should be capable of placing strong limits or possibly observing a positive detection signal with this channel.  Also, if the inelastic splitting isn't too large ($\delta \lsim 150$ keV) then there is also good potential for observing inelastic scattering, which could be discriminated against elastic scattering by the shape of the event spectrum, thus giving a unique window onto the size of the inelastic splitting
among sneutrino states.

Therefore upcoming direct and indirect detection experiments will provide a complimentary test of the validity of this model of DM.

\subsection{Collider Signatures}

Due to the small mass splittings of the six lightest sneutrino states, to a good approximation all six states would be produced in equal multiplicity at the LHC, making a determination of flavour structures at the LHC very difficult.  However it may be possible to identify sneutrinos as the LSP.  The LHC signatures of mixed sneutrinos have been studied previously \cite{Thomas:2007bu} where it was found that mixed sneutrinos can be distinguished from the MSSM over large regions of parameter space.   If the NLSP is a right-handed slepton then large lepton multiplicities may result from the decays of more massive sparticles.  In particular decays of the lightest neutralino lead to opposite-sign, same-flavour dilepton signatures.  This can be distinguished from similar MSSM scenarios by observing the shape of the dilepton invariant mass distribution \cite{Thomas:2007bu}.
For large $\sin\theta$ decay chains starting with squarks can lead to jet-lepton signatures which are present in the MSSM, but could be distinguished by observing sneutrino-charged slepton mass splittings which arise only through electroweak symmetry breaking in the MSSM
\cite{Thomas:2007bu}.

Finally, if the sneutrinos were to lie in the $Z$-funnel region, $M_{\tilde{\nu}_L} \sim 45$ GeV, as in the left panel of Figure \ref{ml200&bf}, Higgs searches at the LHC and ILC could be dramatically altered because the large $A$-term implies that the dominant decay of the Higgs will be to two sneutrinos. For the parameters of Figure \ref{ml200&bf} and $\sin\theta=0.2$, the Higgs decay width $\Gamma_h$ will increase by an order of magnitude from $\Gamma_h\sim 5$ MeV to $\Gamma_h\sim 50$ MeV. Although not directly measureable at the ILC, this could be determined indirectly from $\Gamma_h=\Gamma(H\rightarrow WW^*)/Br(H\rightarrow WW^*)$ where $Br(H\rightarrow WW^*)$ is directly measured and $\Gamma(H\rightarrow WW^*)$ is estimated from the measurement of the $WWH$ coupling \cite{Yamamoto:2007ig}. The branching fractions for the decays to SM particles, in particular, $b\bar{b}$ and $\gamma\gamma$ would also be lower by an order of magnitude, making Higgs searches at the LHC more difficult.

Therefore, it seems that this model of mixed sneutrino DM has signals at colliders, and is testable through direct and indirect detection experiments over much of the parameter space.

\section{Conclusions}\label{conclusions}

The model detailed herein constitutes a viable model of both DM and neutrino mass generation in which the DM is comprised of mixed lhd-rhd sneutrinos in the same flavour mixtures as the neutrino mass eigenstates.   We have shown in detail that it is possible that the observed
neutrino masses and mixings arise as a consequence of supersymmetry breaking effects in the sneutrino sector, consistent with all
experimental constraints.   The prospects for indirect detection of the associate sneutrino DM from solar capture are good, owing to the large sneutrino-nucleon cross-section afforded by the inelastic splitting of Z-coupled states as shown in Figures \ref{Limits}, \ref{ml200&bf}, and \ref{VaryDe}.  For some regions of the allowed parameter space a potential `smoking gun' signature of this model would be monochromatic, non-oscillating, neutrinos comprised of just one mass eigenstate which originate from sneutrino annihilation in the Sun, see Figure \ref{spectrum}.  The current generation of direct detection experiments place strong limits at low sneutrino masses and at high mass if the inelastic splitting is small, and next generation experiments can explore much of the parameter space, see Figures \ref{Limits}, \ref{ml200&bf}, and \ref{VaryDe}.  There is a prospect of seeing both elastic and inelastic scattering in these experiments, allowing
the size of the sneutrino splitting $\delta_{\alpha, L}$ to be directly measured.   Despite being one of the earliest examples of inelastic DM,
limits from solar capture completely exclude this model as an explanation of DAMA/LIBRA.   The mixed lhd-rhd sneutrinos can have
signatures at the LHC, and can change aspects of Higgs physics.
Finally, we have shown that a simple extension of the MSSM leads to a much richer structure in the DM sector linking different areas of beyond-the-Standard-Model physics, namely dark matter and neutrino physics.

\section{Acknowledgements}
We gratefully thank Markus Ahlers, John Beacom, Lawrence Hall, Philipp Mertsch, Subir Sarkar and Stephen West for conversations and Itay Yavin for conversations and for data relating to solar capture limits.  CM and MM are supported by STFC Postgraduate Studentships.  We also acknowledge support by the EU Marie Curie Network ÒUniverseNetÓ (HPRN-CT-2006-035863), and JMR by a Royal Society Wolfson Merit Award.

\appendix
\section{Sneutrino Masses and Mixings}\label{A}
After electroweak symmetry breaking the sneutrino mass matrix has the form:
\begin{equation}
M^2_{ij} = \left( \begin{array}{cccc}
M^2_L \mathbbm{1}_3 & A v \sin \beta   \mathbbm{1}_3 & 0 & \lambda_{ij} M_N v \sin \beta \\
A v \sin \beta   \mathbbm{1}_3 & M^2_R  \mathbbm{1}_3 & \lambda_{ij} M_N v \sin \beta & \lambda_{ij} M^2_{B} \\
0 & \lambda_{ij} M_N v \sin \beta & M^2_L  \mathbbm{1}_3 & A v \sin \beta   \mathbbm{1}_3 \\
\lambda_{ij} M_N v \sin \beta & \lambda_{ij} M^2_{B} & A v \sin \beta   \mathbbm{1}_3 & M^2_R  \mathbbm{1}_3 \end{array} \right)
\end{equation}
where the basis is $(\tilde{\boldsymbol \nu}^*,\tilde{\boldsymbol n},\tilde{\boldsymbol \nu},\tilde{\boldsymbol n}^*)$ and individual terms are defined in Section~\ref{sneumass}.  If $U_{\lambda}$ is the matrix that diagonalises the $\lambda$ matrix as; $U_{\lambda} \cdot \lambda \cdot U^{\dagger}_{\lambda} = Diag(\lambda_1,\lambda_2,\lambda_3)$, and $P_6$ is a permutation matrix given by;
\begin{equation}\label{p6}
P_6 = \left( \begin{array}{cccccc}
1 & 0 & 0 & 0 & 0 & 0 \\
0 & 0 & 0 & 1 & 0 & 0 \\
0 & 1 & 0 & 0 & 0 & 0 \\
0 & 0 & 0 & 0 & 1 & 0 \\
0 & 0 & 1 & 0 & 0 & 0 \\
0 & 0 & 0 & 0 & 0 & 1 \end{array} \right)
\end{equation}
which rotates $6\times6$ matrices made up of four diagonal $3\times3$ blocks down to three $2\times2$ matrices along the diagonal, then the product;
\begin{equation}\label{mixmat}
U = \frac{1}{\sqrt{2}}
\left( \begin{array}{cc}
P_{6} & 0 \\
0 & P_{6} \end{array} \right)
\cdot
\left( \begin{array}{cc}
\mathbbm{1}_6 & -\mathbbm{1}_6 \\
i \mathbbm{1}_6 & i \mathbbm{1}_6 \end{array} \right)
\cdot
\left( \begin{array}{cc}
\mathbbm{1}_2 \otimes U_{\lambda} & 0 \\
0 & \mathbbm{1}_2 \otimes U_{\lambda} \end{array} \right)
\end{equation}
rotates the sneutrino mass matrix down to the block-diagonal form of six $2\times2$ matrices described in Section~\ref{sneumass}.

\section{Neutrino Masses and Mixings}\label{B}
From the form of the neutrino mass formula one can see that the neutrino mass matrix is diagonalized by the matrix $U_{\lambda}$ to all orders.  Considering just the mixing part of eq.(\ref{loopmass}) we have the term:
\begin{equation}\label{loopmass2}
{M_{\nu_{ij}}} \propto \sum^{12}_{\alpha=1}   U_{i,\alpha}^{\dagger} L(r_\alpha,m_{\chi_x}) U_{\alpha,j+6}
\end{equation}
$L(r_\alpha,m_{\chi_x})$ is a diagonal matrix and, as $P_6$ is a permutation matrix, $P^{\dagger}_6 \cdot Diag(6\times6) \cdot P_6$ is also diagonal.  Therefore the only non-diagonal contribution to the neutrino mass matrix is from the $U_{\lambda}$ part of the sneutrino mixing matrix and $U_{\lambda} \cdot M_\nu \cdot U^{\dagger}_{\lambda}$ must be diagonal.

\subsection{Perturbative Mass Formula}
Here we give the full form of eq.(\ref{pertmass}).  Defining $x_x = M_{\tilde{\nu}_L}/m_{\chi_x}$ and $y_x = M_{\tilde{\nu}_H}/m_{\chi_x}$, then:
\begin{equation}\label{pertmass2}\begin{split}
{M_{\nu_{ij}}} &= \lambda_{ij} \left( \frac{M_Z M_B}{4 \pi v} \right)^2 \sum^4_{x=1} \frac{1}{M_{\chi_x}} (N_{x1} \sin \theta_W - N_{x2} \cos \theta_W)^2 \\
 &\quad \cdot \left(\sin^2\theta(\cos^2\theta-a_\nu \sin 2\theta)L_1(x_x,y_x)+\cos^2\theta(\sin^2\theta+a_\nu \sin 2\theta)L_2(x_x,y_x)\right)
 \end{split}\end{equation}
where
\begin{equation}
L_1(x,y)=\frac{1}{x^2-y^2}\left( \frac{x^2 \ln x^2}{1-x^2}-\frac{y^2 \ln y^2}{1-y^2} \right)-\frac{1}{1-y^2}\left(1+\frac{\ln y^2}{1-y^2} \right)
\end{equation}
and
\begin{equation}
L_2(x,y)=\frac{1}{x^2-y^2}\left( \frac{x^2 \ln x^2}{1-x^2}-\frac{y^2 \ln y^2}{1-y^2} \right)-\frac{1}{1-x^2}\left(1+\frac{\ln x^2}{1-x^2} \right)
\end{equation}

\bibliographystyle{JHEP}
\bibliography{sneurefs}

\end{document}